\documentclass[twocolumn,showpacs,preprintnumbers,amsmath,amssymb]{revtex4}
\usepackage{epsfig}
\usepackage{dcolumn}
\usepackage{bm}
\DeclareGraphicsExtensions{.eps,.ps,.eps.gz,.ps.gz,.eps.Z,{}}

\newcommand{\vx}{{\bf x}}

\newcommand{\vq}{{\bf q}}

\newcommand{\vt}{{\bf t}}

\newcommand{\vk}{{\bf k}}

\newcommand{\dd}{{\rm d}}

\newcommand{\mg}{\big<}
\newcommand{\md}{\big>}

\newcommand{\mC}{{\cal C}}

\newcommand{\mF}{{\cal F}}

\newcommand{\mL}{{\cal L}}
\newcommand{\mM}{{\cal M}}
\newcommand{\mI}{{\cal I}}

\newcommand{\mW}{{\cal W}}

\newcommand{\mT}{{\cal T}}

\newcommand{\Dirac}{\delta_{\rm D}}

\newcommand{\beq}{\begin{equation}}
\newcommand{\eeq}{\end{equation}}
\newcommand{\beqa}{\begin{eqnarray}}
\newcommand{\eeqa}{\end{eqnarray}}
 \def\la{\mathrel{\mathpalette\fun <}}
 
 \def\fun#1#2{\lower3.6pt\vbox{\baselineskip0pt\lineskip.9pt
        \ialign{$\mathsurround=0pt#1\hfill##\hfil$\crcr#2\crcr\sim\crcr}}}

\def\kvecMpc{\, h \, {\rm Mpc}^{-1}}

\begin{document}

\title{Multi-Point Propagators in Cosmological Gravitational Instability}
\author{Francis Bernardeau}
\affiliation{Institut de Physique Th{\'e}orique,
             CEA, IPhT,
             CNRS, URA 2306, F-91191 Gif-sur-Yvette, France
	     {\rm and} Canadian Institute for Theoretical Astrophysics, University of Toronto,
60 St. George street, Toronto, Ontario M5S 3H8, Canada}
\author{Mart\'{\i}n Crocce}
\affiliation{Institut de Ci\`encies de l'Espai, IEEC-CSIC, Campus UAB,
Facultat de Ci\`encies, Torre C5 par-2,  Barcelona 08193, Spain}
\author{Rom\'an Scoccimarro}
\affiliation{Center for Cosmology and Particle Physics, Department of Physics,  
New York University, New York, NY 10003}
\vspace{.2 cm}
\date{\today}
\vspace{.2 cm}
\begin{abstract}

We introduce the concept of multi-point propagators between linear cosmic fields and their nonlinear counterparts in the context of cosmological perturbation theory.  Such functions express how a non-linearly evolved Fourier mode depends on the full ensemble of modes in the initial density field. We identify and resum the dominant diagrams in the large-$k$ limit, showing explicitly that multi-point propagators decay into the nonlinear regime at the same rate as the two-point propagator. These analytic results generalize the large-$k$ limit behavior of the two-point propagator to arbitrary order. We measure the three-point propagator  as a function of triangle shape in numerical simulations and confirm the results of our high-$k$ resummation.  We show that any $n-$point spectrum can be reconstructed from multi-point propagators, which leads to a physical connection between nonlinear corrections to the power spectrum at small scales and higher-order correlations at large scales. As a first application of these results, we calculate the reduced bispectrum at one-loop in renormalized perturbation theory and show that we can predict the decrease in its dependence on triangle shape at redshift zero, when standard perturbation theory is least successful. 

\end{abstract}

 \vskip2pc

\maketitle
\section{Introduction}

Detailed understanding of the development of gravitational instability  is one of the central issues for the study of structure formation in observational cosmology. Although the details of the formation of objects are affected at small scales by the presence of baryonic matter, in the context of a dark matter dominated universe it is thought that the global properties of the large-scale matter distribution is determined to a large extent by the collisionless dynamics of dark matter. Therefore, the Vlasov equation, i.e. the collisionless limit of the Boltzmann equation, describes the dynamics we are interested in (see e.g.~\cite{1980lssu.book.....P,2002PhR...367....1B} for details). This is for instance what pure N-body cosmological simulations attempt to solve. At large scales, where orbit crossing is insignificant, the Vlasov equation reduces to the dynamics a pressureless perfect fluid. 

Although the problem is well posed, its resolution beyond the linear regime is still largely an open question. Some ideas have been developed in the literature (see e.g.~\cite{2002PhR...367....1B} and references therein) but to make predictions beyond the linear regime for  the density power spectrum, the standard approach in the literature has been to resort to semi-analytic prescriptions obtained by fitting to numerical simulations. These fitting formulae, e.g.~\cite{1996MNRAS.280L..19P,2003MNRAS.341.1311S}, originate either from the near universal transform advocated in \cite{1991ApJ...374L...1H} or are based on an even more empirical construction, the halo model~\cite{2002PhR...372....1C}. 

These prescriptions however offer predictions for the power spectrum with relatively low accuracy, at the level of 5-10\%, and are insecure in cases of non standard cosmological models. With the advent of precision observations of the large-scale structure, in particular in the context of the search for dark energy signatures in the growth of structure, this is clearly insufficient~\cite{2005APh....23..369H}. It then becomes crucial to build reliable, and well controlled, prescriptions for describing the nonlinear evolution of the power spectrum and higher order correlation functions such as the bispectrum. For the latter there is not much published work on the expected behavior beyond tree order, although fitting formulae analogous to the power spectrum case have been proposed~\cite{2001MNRAS.325.1312S,2005ApJ...619..667H,2007arXiv0707.1594P}. Recently, a promising new approach to fitting measurements of statistics in N-body simulations and interpolating to other cosmologies has been developed in~\cite{2007astro.ph..2348H}. 
 
With the Renormalized Perturbation Theory (hereafter RPT) formalism introduced in \cite{2006PhRvD..73f3519C},  a new approach emerged for the construction of perturbation schemes applicable to such  observables. It has been successfully applied to the two-point propagator~\cite{2006PhRvD..73f3520C}, and to the density power spectrum in the weakly non-linear regime~\cite{2008PhRvD..77b3533C}. These results show in particular how the acoustic peaks are affected by the mode-coupling effects in the growth of structure. This reformulation of the gravitational perturbation theory also finds alternative approaches~\cite{2007PhRvD..75d3514M,2007A&A...465..725V,2007JCAP...06..026M,2008ApJ...674..617T,2007arXiv0711.2521M}, which give further insights into resummation schemes introduced by RPT.

The aim of this article is to show that some of the core results that have been originally discovered for the two-point propagator in~\cite{2006PhRvD..73f3520C} can also be obtained in the case of multi-point propagators, thus extending exact known results regarding the evolution of cosmic gravitational instability.

The plan of the paper is the following. We first succinctly recall the general formalism and the results that have already been obtained at the two-point level~\cite{2006PhRvD..73f3520C}, involving the cross-correlation between fields at the initial and final conditions. As shown in previous studies, the behavior of this correlator at large-$k$ is of crucial importance and provides us with solid ground for the expansion of the correlation functions of interest~\cite{2008PhRvD..77b3533C}. We then show that the three-point propagator, namely the cross-correlation between the final density field and two of the initial modes, can also be computed explicitly in the high-$k$ limit. This quantity reflects how the final density is sensitive to the two mode-mode couplings. We then extend this result to propagators of arbitrary number of points. This is the central result of this paper. 

To contrast our results with numerical simulations, we extend the method developed in~\cite{2006PhRvD..73f3520C} to measure multi-point propagators. This allows us to compute the three-point propagator in numerical simulations against our predictions in the large-$k$ regime. We finally comment on the relevance of these results for the construction of power spectrum and bispectrum resummation schemes, and provide a first application to the case of the bispectrum.

\section{Dynamics}
\label{nonlineardynamics}

Cosmological gravitational instability  in a pressureless fluid in the single stream limit (when orbit crossing can be neglected) is governed by a set of three coupled equations relating the density contrast $\delta=\rho/\bar \rho - 1$, the peculiar velocity field ${\bf v}$ and the gravitational potential. Consistently with this limit one can assume the fluid to be irrotational and thus fully described by $\delta$ and the velocity divergence, $\theta\equiv\nabla\cdot{\bf v}$. Following~\cite{1998MNRAS.299.1097S}, the equations of motion in Fourier space can be written in a compact form with the use of the two component quantity
\begin{equation}
\Psi_a(\vk,s) \equiv \Big( \delta(\vk,s),\ -\theta(\vk,s)/{\cal H} \Big),
\label{2vector}
\end{equation}
where the index $a=1,2$ selects the density or velocity components, ${\cal H}\equiv {d\ln a/{d\tau}}$  is the conformal expansion rate, $a(\tau)$ the cosmological scale factor and $\tau$ conformal time. The time variable $s$ is defined from the scale factor by 
\begin{equation}
s\equiv\ln a(\tau),
\label{timevariable}
\end{equation}
and corresponds to the number of e-folds of expansion. We will first consider a cosmology with $\Omega_m=1$ and $\Omega_{\Lambda}=0$. The extension to a general case is obtained at the scales we are interested in to a very good approximation  by replacing $a(\tau)$ by the linear growth factor of the cosmology under consideration \cite{2002PhR...367....1B}. 

The equations of motion then read (we henceforth use the convention that  repeated Fourier arguments are integrated over),
\begin{eqnarray}
\partial_{s} \Psi_a(\vk,s) + \Omega_{ab} \Psi_b(\vk,s) &=&\nonumber\\
&&\hspace{-2cm} \gamma_{abc}(\vk,\vk_1,\vk_2) \ \Psi_b(\vk_1,s) \ \Psi_c(\vk_2,s), \ \ \ \
\label{eom}
\end{eqnarray}
where 
\begin{equation}
\Omega_{ab} \equiv \Bigg[ 
\begin{array}{cc}
0 & -1 \\ -3/2 & 1/2 
\end{array}        \Bigg],
\end{equation}
and the {\em symmetrized vertex} matrix $\gamma_{abc}$ describes the non linear interactions between different Fourier modes and is given by
\begin{eqnarray}
\gamma_{222}(\vk,\vk_1,\vk_2)&=&\Dirac(\vk-\vk_1-\vk_2) \ {|\vk_1+\vk_2|^2 (\vk_1
\cdot\vk_2 )\over{2 k_1^2 k_2^2}}, \nonumber \\
\gamma_{121}(\vk,\vk_1,\vk_2)&=&\Dirac(\vk-\vk_1-\vk_2) \  {(\vk_1+\vk_2) \cdot
\vk_1\over{2 k_1^2}},
\label{vertexdefinition}
\end{eqnarray}
$\gamma_{112}(\vk,\vk_1,\vk_2)=\gamma_{121}(\vk,\vk_2,\vk_1)$, and $\gamma=0$ otherwise; with $\Dirac$ denoting the Dirac delta distribution. 
The formal integral solution to Eq.~(\ref{eom}) is given by (see \cite{1998MNRAS.299.1097S,2001NYASA.927...13S,2006PhRvD..73f3519C} for a detailed derivation), 
\begin{eqnarray}
\Psi_a(\vk,s) &=& g_{ab}(s) \ \phi_b(\vk) +  \int_0^s  \dd s' \ g_{ab}(s-s') \nonumber \\
&& \times \gamma_{bcd}^{(\rm s)}(\vk,\vk_1,\vk_2) \Psi_c(\vk_1,s') \Psi_d(\vk_2,s'),
\label{eomi}
\end{eqnarray}
where  $\phi_a(\vk)\equiv\Psi_a(\vk,s=0)$ denotes the initial conditions, set when the linear growth factor $a(\tau)=1$ and $s=0$. The {\em linear propagator} $g_{ab}(s)$ is given by 
\begin{equation}
g_{ab}(s) = \frac{{\rm e}^{s}}{5}
\Bigg[ \begin{array}{rr} 3 & 2 \\ 3 & 2 \end{array} \Bigg] -
\frac{{\rm e}^{-3s/2}}{5}
\Bigg[ \begin{array}{rr} -2 & 2 \\ 3 & -3 \end{array} \Bigg],
\label{prop}
\end{equation} 
for $s\geq 0$, whereas $g_{ab}(s) =0$ for $s<0$ due to causality, and $g_{ab}(s) \rightarrow \delta_{ab}$ as $s\rightarrow 0^{+}$. We will work with  {\it growing mode initial conditions}, as usually assumed in numerical simulations, for which
\begin{equation}
\phi_a(\vk) = u_a \, \delta_0(\vk),
\label{gmic}
\end{equation}
and $u_a=(1,1)$. 

A perturbative solution to Eq.~(\ref{eomi}) can be obtained expanding the fields in terms of the initial ones,
\begin{equation}
\Psi_{a}(\vk,s)=\sum_{n=1}^{\infty}\Psi_{a}^{(n)}(\vk,s),
\label{PsiExpansion}
\end{equation}
with
\begin{eqnarray}
\Psi_{a}^{(n)}(\vk,s)=\int\dd^3\vk_{1}\dots\dd^3\vk_{n}\ 
\Dirac(\vk-\vk_{1\dots n}) \nonumber \\
\times \,\mF^{(n)}_{a b_1 b_2 \ldots b_n}(\vk_{1},\dots,\vk_{n};s)
\phi_{b_1}(\vk_{1})\dots\phi_{b_n}(\vk_{n}),
\label{mFndef}
\end{eqnarray}
where $\mF^{(n)}$ are fully symmetric functions of the wave-vectors that can be obtained recursively in terms of $g_{ab}$ and $\gamma_{abc}$ \cite{2002PhR...367....1B}.  Note that these functions have a non-trivial time dependence because they also include sub-leading terms in ${\rm e}^s$. Their fastest growing contribution is of course given by the well known $\{F_n,G_n\}$ kernels in PT (assuming growing mode initial conditions),
\begin{equation}
\mF^{(n)}_a=\exp(ns)\ \{F_n(\vk_1,..,\vk_n),G_n(\vk_1,..,\vk_n)\} 
\end{equation}
for $a=1,2$ (density or velocity divergence fields respectively).

\begin{figure}
\centerline{\epsfig {figure=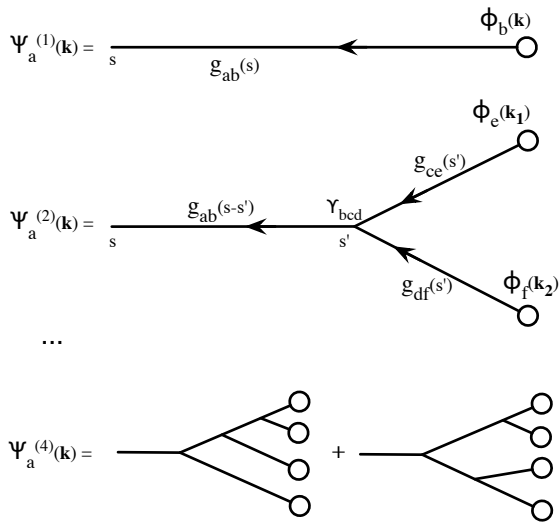,width=8cm}}
\caption{Diagrammatic representation of the series expansion of $\Psi_{a}(\vk)$ up to fourth order in the initial conditions $\phi$. Time increases along each segment according to the arrow and each segment bears a factor $g_{cd}(s_{f}-s_{i})$ if $s_{i}$ is the initial time and $s_{f}$ is the final time. 
At each initial point and each vertex point there is a sum over the component indices; a sum over the incoming wave modes is also implicit and, finally, the time coordinate of the vertex points is integrated from $s=0$ to the final time $s$ according to the time ordering of each diagram. For instance, at fourth order there are two different possible topologies.}
\label{DiagramPsiExpansion}
\end{figure}

The relation in Eq.~(\ref{eomi}) can be used to construct a diagrammatic representation of the fields $\Psi^{(n)}$. As an illustration, we show in Fig.~\ref{DiagramPsiExpansion} diagrams corresponding to the local field expansion up to fourth order.

A detailed description of the procedure to draw the diagrams and compute their values can be found in \cite{2006PhRvD..73f3519C}, we can briefly summarize these rules here as follows. In Fig. \ref{DiagramPsiExpansion} the open circles represent the initial conditions $\phi_b(\vk)$, where $b=1$ ($b=2$) corresponds to the density (velocity divergence) field, and the line emerging from it carries a wavenumber $\vk$. Lines are time-oriented (with time direction represented by an arrow) and have different indices at both ends, say $a$ and $b$. Each line represents linear evolution described by the propagator $g_{ab}(s_{f}-s_{i})$ from time $s_{i}$ to time $s_{f}$. Each nonlinear interaction between modes is represented by a vertex, which due to quadratic nonlinearities in the equations of motion is the convergence point of necessarily two incoming lines, with wavenumber say $\vq_{1}$ and $\vq_{2}$, and one outgoing line with wavenumber $\vq=\vq_1+\vq_2$. Each vertex in a diagram then represents the matrix $\gamma_{abc}(\vq,\vq_{1},\vq_{2})$. It is further understood  in Fig.~\ref{DiagramPsiExpansion} that internal indices are summed over and interaction times are integrated out in the full interval $[0,s]$. 

Note that at the level of the field $\Psi$, all diagrams are tree diagrams with no loops. This is unlike in standard field theories, and is due to the lack of  ``antiparticles" (both in the strict sense of the word, like in quantum field theory, and at the quasiparticle level, as e.g. holes in condensed matter systems). However, loop diagrams appear once we calculate statistical averages such as correlators between fields. Other unusual properties compared to standard field theories are the non-trivial $k$-dependence of vertices (see also~ \cite{2008JCAP...01..029R}) and last, but not least, a strong breaking of time-translation invariance. In fact, the theory is unstable with perturbations growing as a function of time from their initial conditions (represented by density and velocity fields after decoupling). 

Having written the final conditions of interest $\Psi$ in terms of initial conditions $\phi$, to build diagrammatic representations for statistical quantities such as the nonlinear propagator or the power spectrum, one needs to specify the statistical properties of the initial conditions. Here we will assume Gaussian initial conditions, in which case the statistical properties are encoded in the initial spectrum of fluctuations,
\begin{equation}
\mg\phi_{a}(\vq)\phi_{b}(\vq')\md=\Dirac(\vq+\vq')  P_{ab}(q). 
\label{Spectra}
\end{equation} 
with $P_{ab}=u_a u_b P_0$ for growing-mode initial conditions, i.e. if Eq.~(\ref{gmic}) is satisfied. Ensemble averages are therefore obtained by gluing together the open circles in the diagrams by pairs (according to the calculation rules for Gaussian fields) to form a symbol $\otimes$, which will appear in all diagrams in the figures below. This symbol represents the initial power spectrum $P_{ab}(q)$, when the outgoing lines away from the initial conditions have respectively indices $a$ and $b$ and carry wavenumbers $\vq$ and $\vq'$. Finally, in all diagrams it is understood that internal wavenumbers (corresponding to loops) when such symbols appear are integrated over.

\section{Multi-Point Propagators}

\subsection{The Two-Point Propagator}

The idea at the heart of  RPT is to realize that the diagrammatic series described above for correlation functions can be partially resummed. This amounts to generalizing the operator $g_{ab}$, encoding the linear evolution, to a fully nonlinear one, denoted by $G_{ab}$, that effectively describes time evolution when coupling between modes is present. This generalized operator, called the nonlinear propagator, is formally defined as,
\begin{equation}
\mg\frac{\delta \Psi_{a}(\vk,s)}{\delta
\phi_{b}(\vk')}\md=\Dirac{\left(\vk-\vk'\right)}\,G_{ab}(k,s).
\label{Gabdef}
\end{equation} 
This quantity basically measures  how linear is the transition from initial to final fields. Note that, although certainly $\Psi_{a}(\vk)$ depends on $\phi_{b}(\vk')$ for $\vk'\neq \vk$, translation invariance demands that {\em the expectation value} of the derivative in Eq.~(\ref{Gabdef}) is only non-zero if $\vk=\vk'$~\footnote{Formally, a translation $\vx \rightarrow \vx+\vt$ changes Fourier coefficients by a phase ${\rm e}^{i\vk\cdot\vt}$, which induces a factor ${\rm e}^{i(\vk-\vk')\cdot\vt}$ in Eq.~(\ref{Gabdef}) and hence invariance demands $\vk=\vk'$. A more pedestrian demonstration is to recall that an arbitrary nonlinear contribution to  $\Psi(\vk)$ of, say, $(n+1)$-th order in $\phi$ with momenta $\vq_i$ is restricted by translation invariance to obey $\sum_{i=1}^{n+1}\vq_i=\vk$. Upon taking the derivative in  Eq.~(\ref{Gabdef}) this constraint reads e.g. $\sum_{i=1}^{n}\vq_i+\vk'=\vk$, thus taking expectation values for the $n$ remaining $\phi$'s imposes $\sum_{i=1}^{n}\vq_i=0$, again by translation invariance.}.

Due to nonlinear evolution, one expects $G_{ab}$ to decay to zero at small scales where non-linear effects become important and make $\Psi(\vk)$ very different from a linear transformation of $\phi_{b}(\vk)$. In fact, for Gaussian initial conditions it is possible to give an alternative expression for $G_{ab}$, as a cross-correlation  between final $\Psi_{a}$ and initial $\phi_{b}$ fields~\cite{2006PhRvD..73f3520C},
\begin{equation}
\mg\Psi_{a}(\vk,s)\,\phi_{b}(\vk')\md=G_{ac}(k,s)\,\mg\phi_{c}(\vk)\,\phi_{b}(\vk')\md.
\label{Gabdef2}
\end{equation}
This equivalence gives rise to the alternative name of {\em two-point} propagator and reinforces the idea that decorrelation of fields $\Psi_{a}$ with respect to $\phi_{b}$ mandates that $G_{ab}$ must vanish at small scales. Indeed, it was shown in~\cite{2006PhRvD..73f3520C} that $G_{ab}$ can be computed in the large-$k$ limit and it follows a Gaussian decay with $k$. This was also checked in numerical simulations (see  Fig.~\ref{LargeKlimit} and discussion in sec.~\ref{largekGammap} below). At low-$k$, on the other hand, linear theory is recovered and $G_{ab}(k\rightarrow 0)=g_{ab}$, as expected.
 

Introducing the concept of a two-point propagator helps to understand nonlinear corrections to the power spectrum. The power spectrum can be written as~\cite{2006PhRvD..73f3519C}

\beq
P(k,\tau) = G^2(k,\tau)\, P_0(k) + P_{\rm MC}(k,\tau),
\label{PRPT}
\eeq
where we have explicitly included the time dependence, $P_0$ is the initial density power spectrum, and $P_{\rm MC}$ is the mode-coupling power spectrum. Note that what we call $G$ in this equation is the density propagator $G\equiv G_{11}+G_{12}=G_{1b}u_b$, since we assume growing-mode initial conditions; a completely equivalent equation holds for velocities. The first term describes {\em all} the contributions to $P(k)$ that are proportional to the initial power spectrum at the {\em same scale} $k$, while the second, denotes the contribution to $P(k)$ from modes other than $k$ in the initial spectrum (hence called ``mode-coupling"), i.e. $P_{\rm MC}(k)$ is nonzero even if $P_0$ vanishes at wavenumber $k$ (as long as $P_0$ is not zero everywhere).  

The first term in Eq.~(\ref{PRPT}) has the most direct information of the initial power spectrum, since it is proportional to $P_0$ at the same scale, thus varying $k$ one can probe $P_0(k)$ as long as the first term is dominant, i.e. at large scales where $G$ has not yet decayed significantly. The two-point propagator $G$ can be given an alternative interpretation by noting that Eq.~(\ref{Gabdef2}) says that $(G/D_+)^2$, where $D_+$ is the linear growth factor, is the fraction of the power that grows by linear theory at a given $k$. As $G$ decays, a given mode looses memory about its primordial (linear) value, and an important fraction of the power is contributed by nonlinear effects alone, i.e. $P_{\rm MC}$. However, that does not mean that information on the initial power spectrum is completely lost at such scales. In fact, RPT gives a precise expression for $P_{\rm MC}$ in terms of convolutions over the initial spectrum~(see Fig.~5 in~\cite{2008PhRvD..77b3533C} for how well this works when compared to simulations). However the information on $P_0$ contained in $P_{\rm MC}$ is rather crude, since $P_{\rm MC}(k)$ depends on nonlinear combinations of $P_0(q)$'s  from a wide range of scales $q$ weighted by complicated convolution kernels. 

As we will show in the rest of the paper, these kernels are the (square of) multi-point propagators and, in fact, they also determine the tree-level higher-order spectra (bispectrum, trispectrum, etc). Therefore, multi-point propagators will lead us to a different way of writing the mode-coupling contributions to $P(k)$ (and higher-order spectra), and a direct physical connection between nonlinear corrections to the power spectrum at small scales and higher-order correlations at large scales.

\subsection{Definition of Multi-Point Propagators}
\label{thenonlinearpropagators}

Having discussed the importance of the two-point propagator let us introduce the concept of multi-point propagators, denoted $\Gamma^{(p)}$, as a natural extension of the definition in Eq.~(\ref{Gabdef}),
\begin{equation}
\frac{1}{2}\, \mg\frac{\delta^2 \Psi_{a}(\vk,s)}{\delta\phi_{b}(\vk_{1})\delta\phi_{c}(\vk_{2})}\md
=\Dirac(\vk-\vk_{12})\ \Gamma^{(2)}_{abc}\left(\vk_{1},\vk_{2},s\right)
\label{GammaabcDef}
\end{equation}
for the three-point case and,
\begin{eqnarray}
\frac{1}{p!}\, 
\mg\frac{\delta^p
\Psi_{a}(\vk,s)}{\delta\phi_{b_{1}}(\vk_{1})\dots\delta\phi_{b_{p}}(\vk_{p})}\md
&=&\nonumber\\
&&\hspace{-3cm}\Dirac(\vk-\vk_{1 \ldots p})\ \Gamma^{(p)}_{ab_{1}\dots
b_{p}}\left(\vk_{1},\dots,\vk_{p},s \right),
\label{GammaAllDef}
\end{eqnarray}
where $\vk_{1\ldots p}=\vk_1+\ldots+\vk_p$, for an arbitrary number of points. Note that $\Gamma^{(p)}$ denotes the {\em $(p+1)$-point propagator}, by translation invariance it depends only on $p$ wavenumbers in Fourier space.

The question we want to address in this section is the calculation of $\Gamma^{(p)}$. We will shortly show that exact results can indeed be obtained for $\Gamma^{(p)}$ in the large $k$ limit. This is a nontrivial generalization of the two-point case and constitutes a core result in this paper. In addition, as anticipated above, we will show in section~\ref{recon} that correlation functions can be reconstructed solely in terms of $\Gamma$ functions and initial power spectra $P_0$, e.g. for the power spectrum we will show that,

\begin{eqnarray}
P(k)=\sum_{r\ge 1} r! \int \delta_{\rm D}(\vk-\vq_{1\ldots r}) \left[\Gamma^{(r)}(\vq_1,\ldots,\vq_r)\right]^2 \nonumber \\ \times \, P_0(q_1)\ldots P_0(q_r) \ \dd^3\vq_1 \ldots \dd^3\vq_r, \ \ \ 
\label{gammaexpansion}
\end{eqnarray}
where $\Gamma^{(1)}$ is the two-point propagator $G$ used in Eq.~(\ref{PRPT}) and the mode-coupling spectrum $P_{\rm MC}$ is described here by the terms with $r\geq 2$. Equation~(\ref{gammaexpansion}) is something already found in~\cite{2006PhRvD..73f3519C} within the Zel'dovich approximation, but extended here to the exact dynamics.

\begin{figure}
\centerline{\epsfig {figure=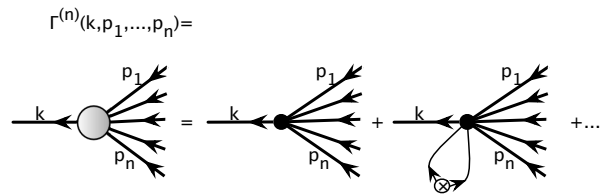,width=8cm}}
\caption{Representation of the first two terms of the multi-point propagator $\Gamma^{(n)}$ in a perturbative expansion. $\Gamma^{(n)}$ represents the average value of the emerging nonlinear mode $\vk$ given $n$ initial modes in the linear regime. Here we show the first two contributions: tree-level and one-loop. Note that each object represents a collection of (topologically) different diagrams.}
\label{Gamma5}
\end{figure}

Formally, the $\Gamma$ functions can be computed order by order in a field expansion and thus they can be represented diagrammatically. Examples of such diagrams are shown in Fig.~\ref{Gamma5}. To lowest order it can be easily shown from Eqs.~(\ref{PsiExpansion},\ref{mFndef},\ref{GammaAllDef}) that $\Gamma^{(p)}$ coincides with the $\mF^{(p)}$ kernels in Eq.~(\ref{mFndef}). Diagramatically this lowest order corresponds to a tree level diagram. Let us explicitly  compute $\Gamma^{(2)}$ starting from the second diagram in Fig.~\ref{DiagramPsiExpansion} and the rules described in Sec.~\ref{nonlineardynamics} to get an expression for $\Psi^{(2)}$, that after the functional derivatives in Eq.~(\ref{GammaAllDef}) gives,
\begin{eqnarray}
\Gamma_{abc,\rm{tree}}^{(2)}(\vk_{1},\vk_{2},s)&=&\int_{0}^{s}\dd s'\ g_{ad}(s-s')
\gamma_{def}(\vk_{1},\vk_{2},\vk)\nonumber \\ && \times g_{eb}(s')\,g_{fc}(s').
\label{Gamma2Tree}
\end{eqnarray}
For the density field ($a=1$) and growing mode initial conditions this leads to,
\begin{eqnarray}
\Gamma^{(2)}_{1bc,{\rm tree}}(\vk_1,\vk_2,s)u_b u_c&=&e^{2s}\left(\frac{5}{7}+\frac{k_1 x}{2 k_2}+\frac{k_2 x}{2 k_1}+\frac{2 x^2}{7}\right) \nonumber \\
&+&e^s\left(-\frac{3}{5}-\frac{k_1 x}{2 k_2}-\frac{k_2 x}{2 k_1}-\frac{2 x^2}{5}\right)\nonumber \\ 
&+&e^{-3s/2}\left(-\frac{4}{35}+\frac{4 x^2}{35}\right),
\label{Gamma1tree}
\end{eqnarray}
with $x=\vk_1\cdot\vk_2/k_1 k_2$. The fastest growing term in Eq.~(\ref{Gamma1tree}) is precisely the $F_2$ kernel in standard PT.

We now show that the equivalence in Eq.~(\ref{Gabdef2}) between two-point propagators and cross-correlations for Gaussian initial conditions, can be extended to multi-point propagators. Let us start with the three-point propagator $\Gamma^{(2)}$. 

On the one hand, an expression for $\Gamma^{(2)}_{abc}\left(\vk_{1},\vk_{2}\right)$ can be formally written in terms of the functions $\mF^{(n)}$ using Eqs.~(\ref{mFndef},\ref{GammaabcDef}). It reads,
\begin{eqnarray}
\Gamma^{(2)}_{abc}\left(\vk_{1},\vk_{2}\right)=
\sum_{p=0}^{\infty}\binom{2p+2}{2p}(2p-1)!!\int\dd^3\vq_{1}\dots\dd^3\vq_{p} 
 \nonumber \\ \times
\ \bar{\mF}_{abc}^{(2p+2)}(\vk_{1},\vk_{2},\vq_{1},-\vq_{1},\dots,-\vq_{p};s)
\,P_{0}(q_{1})\dots P_{0}(q_{p}) \nonumber \\
\label{GammaabcExp2}
\end{eqnarray}
where $\bar{\mF}_{abc}^{(2p+2)}\equiv\mF_{abcd_{1}\dots d_{p}}^{(2p+2)}u_{d_{1}}\dots u_{d_{p}}$, and we assumed Gaussian initial conditions together with Eqs.~(\ref{gmic},\ref{Spectra}). The factor  $\binom{2p+2}{2p}$ is the number of possibilities for choosing $2p$ points out of $2p+2$ and $(2p-1)!!$ is the number of ways $2p$ points can be connected together by pairs.

On the other hand, with the relation (\ref{mFndef}) one can also get a formal expression for the cross-correlation
$\mg\Psi_{a}(\vk)\phi_{b}(-\vk_{1})\phi_{c}(-\vk_{2})\md$. Given the fact that
$\mg\Psi^{(n)}\md\equiv 0$, such an expression involves the same type of contractions of the
$\phi_{a_{i}}(\vk_{i})$ factors leading to, 
\begin{eqnarray}
\mg\Psi_{a}(\vk)\phi_{b}(-\vk_{1})\phi_{c}(-\vk_{2})\md&=&
2\, \Dirac(\vk-\vk_{1}-\vk_{2})\nonumber\\
&&\hspace{-2cm}\times\ \Gamma^{(2)}_{ade}\ P_{db}(k_{1})P_{ec}(k_{2})
\label{ThreeMomentExp}
\end{eqnarray}
with $P_{ab}(k)$ defined in Eq.~(\ref{Spectra}). 
This relation is very useful since it provides a simple way to compute the three-point propagator in numerical simulations out of cross-correlations between final and initial conditions (see Section~\ref{numsimu}).

As expected, the relations in Eqs.~(\ref{GammaabcExp2}-\ref{ThreeMomentExp}) can be straightforwardly extended to any order. We thus have for the generic case,
\begin{eqnarray}
\mg\Psi_{a}(\vk)\phi_{c_{1}}(-\vk_{1})\dots\phi_{c_{p}}(-\vk_{p})\md_{c}&=&
p!\,\Dirac(\vk-\vk_{1\ldots p})\nonumber\\
&&\hspace{-3cm}\times\ \Gamma^{(p)}_{ab_{1}\dots b_{p}}\ P_{b_{1}c_{1}}(k_{1})\dots
P_{b_{p}c_{p}}(k_{p})
\label{npointexpression}
\end{eqnarray}
where $\mg\dots\md_{c}$ stands for the connected part of the ensemble average. 

\section{The large-$k$ behavior of multi-point propagators}
\label{largekGammap}

\begin{figure}
\centerline{\epsfig {figure=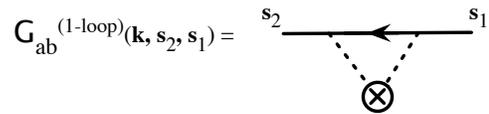,width=7cm}}
\caption{The one-loop contribution to $G_{ab}(k,s_2,s_1)$. The $\otimes$ represents a primordial power spectrum $P_0(q)$ with the corresponding ``loop'' momentum $\vq$ integrated over with weight $(2\pi)^{-3}\int \dd^3 \vq$ . See~\cite{2006PhRvD..73f3520C} for an explicit calculation of this diagram.} 
\label{Gab1loop}
\end{figure}

\subsection{The Large-$k$ Limit of the Two-Point Propagator}

As discussed in the previous section, the two-point propagator $G_{ab}$ generalizes $g_{ab}$ beyond linear theory and thus reflects a key property of the evolved fields. The general properties of $G_{ab}$ have been explored in detail in \cite{2006PhRvD..73f3520C}, but we briefly recall them here to motivate  their generalization to multi-point propagators. 

Following Eqs.~(\ref{PsiExpansion},\ref{mFndef}), and the definition in Eq.~(\ref{Gabdef}), one can expand the function $G_{ab}$ with respect to the amplitude of initial fluctuations,
\begin{equation}
G_{ab}(k,s_{f},s_{i})=g_{ab}(s_{f}-s_{i})+G_{ab}^{\rm{1loop}}(k,s_{f},s_{i})+\dots
\label{GabExpansion}
\end{equation}
where $G_{ab}^{\rm{1loop}}(k,s_{f},s_{i})$ is the first nonlinear correction term, describing the transition into the nonlinear regime. Graphically, this term corresponds to a ``one-loop'' diagram (i.e. an integral over $P_0$), which is shown in Fig.~\ref{Gab1loop}.

\begin{figure}
\centerline{\epsfig {figure=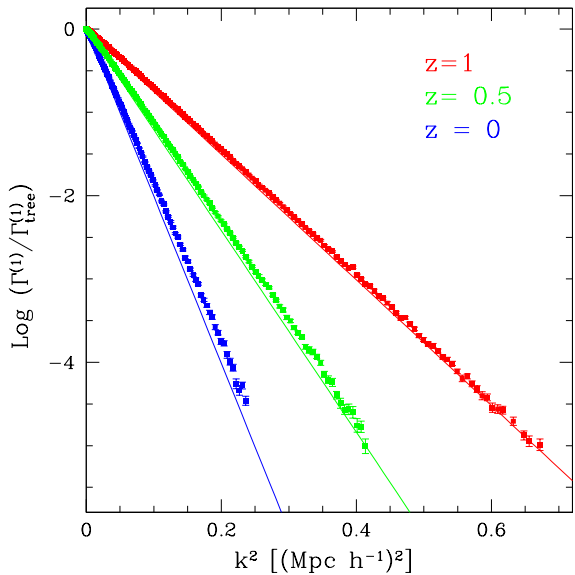,width=8cm}}
\caption{The large-$k$ limit of the two-point density propagator $\Gamma^{(1)}$. Symbols correspond to measurements in numerical simulations at redshifts $z=1,0.5$ and $z=0$ (top to bottom), see text for details. The solid lines correspond to the large-$k$ limit expression given in Eq.~(\ref{GabHighk}). The linear relation obtained by plotting $\log G$ vs. $k^2$ makes evident that the suppression of $G$  is indeed Gaussian in the high-$k$ limit. Moreover, the slope is very well predicted by Eqs.~(\ref{GabHighk},\ref{sigmavdef}). }
\label{LargeKlimit}
\end{figure}

As nonlinear effects become important $G_{ab}$ is expected to decay to zero since they erase the one-to-one correspondence of modes valid in the linear regime. This introduces a characteristic scale that describes the decay length of the two-point propagator. It was shown in~\cite{2006PhRvD..73f3520C} that this decay can be computed exactly in the high-$k$ limit, where a subset of diagrams is expected to provide the dominant contribution. Following a line of calculation that we will use again shortly, it was shown that in the large-$k$ limit,

\begin{equation}
G_{ab}(k,s_{f},s_{i})=\exp\left(-\frac{k^2}{2}\sigma_{v}^2(e^{s_{f}}-e^{s_{i}})^2\right)\,g_{ab}(s_f-s_i),
\label{GabHighk}
\end{equation}
where the characteristic decay length is determined by the {\em rms} velocity fluctuations
\begin{equation}
\sigma_{v}^2={1\over 3} \int_{0}^{\infty}\frac{\dd^3\vk}{k^2}\,P_{0}(k).
\label{sigmavdef}
\end{equation}
In~\cite{2006PhRvD..73f3520C}, it is shown how to match this result valid for $k\sigma_v \gg 1$ to the low-$k$ behavior described by Eq.~(\ref{GabExpansion}), to obtain a prescription for its full time and $k$ dependence. This prescription was found to be in good agreement with numerical simulations at all scales and different redshifts for density and velocity divergence propagators.

Here, we concentrate on the large-$k$ behavior of the density propagator from growing-mode initial conditions, $\Gamma^{(1)}\equiv \Gamma^{(1)}_{1b}u_b=G_{11}+G_{12}$ (we will henceforth use both $G$ and $\Gamma^{(1)}$ to refer to the two-point propagator). We use the algorithm presented in~\cite{2006PhRvD..73f3520C} to measure $\Gamma^{(1)}$ based on the cross-correlation property in Eq.~(\ref{Gabdef2}).  We defer a description of the simulations used here until section~\ref{numsimu} below. Figure~\ref{LargeKlimit} shows $\Gamma^{(1)}$ normalized by the linear growth factor $\Gamma^{(1)}_{\rm tree}=g_{11}+g_{12}$, with $g_{ab}$ the linear propagator defined in Eq.~(\ref{prop}); the unusual notation for the growth factor is used here to emphasize that it is given by the tree-contributions to the two-point propagator, this will have a natural generalization for multi-point propagators.  The figure shows $\log \Gamma^{(1)}$ vs. $\log k^2$ to emphasize the Gaussian decay very well predicted by Eq.~(\ref{GabHighk}) at all redshifts with  characteristic scale given by Eq.~(\ref{sigmavdef}).

In the following sections we investigate similar properties, regimes and measurements to those carried out with $G_{ab}$, to the case of the three-point propagator $\Gamma^{(2)}$ and, when possible, extend this to $\Gamma^{(n)}$.

\subsection{Dominant diagrams and principal trees}

\begin{figure}
\centerline{\epsfig {figure=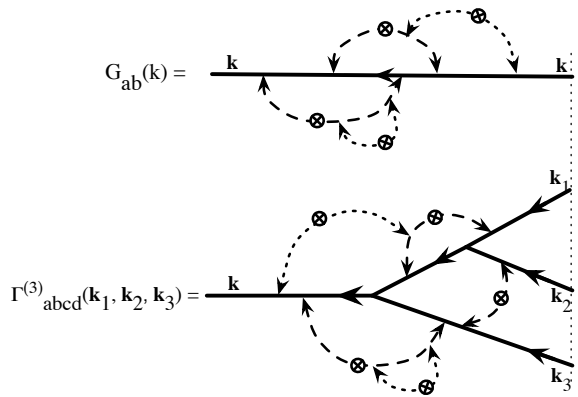,width=8cm}}
\caption{Example of diagrams contributing to $G_{ab}(k)$ (top) and $\Gamma^{(3)}_{abcd}(\vk,\vk_{1},\vk_{2},\vk_{3})$ (bottom). The dominant contribution after resumming all possible configurations is expected to come from those diagrams where all loops are directly connected to the principal line (top) or principal tree (bottom). The principal line and tree are drawn with a thick solid line. The dominant loops are those drawn by dashed lines, while the sub-dominant loops are those in dotted lines.}
\label{PrincipalTrees}
\end{figure}

\begin{figure}[t!]
\centerline{\epsfig {figure=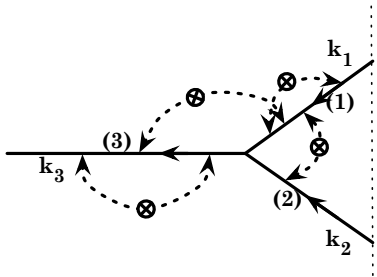,width=5cm}}
\caption{Dominant loop contributions for the three-point propagator  $\Gamma^{(2)}_{abc}(\vk_{1},\vk_{2},\vk_{3})$. The final expression is obtained by the sum of an infinite number of such loops and over all possible interaction times. In the large-$k$ limit these terms can be computed exactly.} 
\label{Gamma2Loops}
\end{figure}

To study the high-$k$ regime of the propagators, the first step is to identify the set of diagrams that is expected to dominate the large-$k$ behavior. This identification is closely linked to the concept of {\em principal line} and its generalization, the {\em principal tree}, that all the diagrams for propagators contain. These concepts are illustrated in the diagrams appearing in Fig.~\ref{PrincipalTrees}. For instance, in the top diagram there is a unique way to go from initial time (symbolized by the vertical dotted line) to the final time without crossing a $\otimes$, thus moving always in the direction of increasing time. It's easy to check that this is true for any diagram contributing to $G_{ab}$. This is what we call its {\em principal line}. 

Similarly there exists a unique tree, the {\em principal tree}, that joins the initial time to the final time in the bottom diagram of Fig.~\ref{PrincipalTrees}. The number of branches of this tree is equal to the number of points minus one for the propagator under consideration (i.e. the number of function derivatives in the definition in Eq.~(\ref{GammaAllDef})). Again, it's easy to check that the existence of a principal tree is general and does not depend on the number of loops that are attached to it, although at a given order different diagrams may have different principal trees (see bottom diagrams of Fig.~\ref{DiagramPsiExpansion}). Principal trees appear then as simple generalization of the concept of principal line initially introduced in \cite{2006PhRvD..73f3520C}.

We now note that in the expansion series, all diagrams can be seen as a principal tree with loops attached to its branches. Following arguments developed in \cite{2006PhRvD..73f3520C}, we claim that the general behavior of the multi-point propagators are dominated in the large-$k$ limit by those diagrams in which every loop is directly connected to the principal tree. 

This idea follows from the fact that multi-point propagators are a measure of the correlation between final and initial configurations, see Eqs.~(\ref{Gabdef2},\ref{ThreeMomentExp},\ref{npointexpression}). And diagrams that contain only such loops are more effective in the process of correlating the final field with the initial conditions (that is, the modes running along the principal tree always interact with growing modes coming from the initial conditions).  In particular, this argument remains valid at large-$k$ where these diagrams can be computed exactly, as we will show in subsequent sections.

As explicit examples, we show in the bottom diagram of Fig.~\ref{PrincipalTrees} in dashed lines the kind of loops that contribute dominantly to the large-$k$ limit of $\Gamma^{(3)}$ (while the principal tree is represented by solid lines and the sub-dominant loops in dotted lines) . In Fig.~\ref{Gamma2Loops} we show some of the dominant loop contributions to $\Gamma^{(2)}$, as they are the subject of next section.

These examples can be used to build a simple picture of the different loop contributions. Let us take the case of the two-point propagator. Its one-loop contribution is shown in Fig.~\ref{Gab1loop}. As shown in~\cite{2006PhRvD..73f3520C}, this contribution is dominated by loop momenta of order $q\simeq \sigma_v^{-1}$, thus is in the high-$k$ limit ($k\sigma_v\gg1$), most of the momentum is flowing along the principal line of the diagram. Now consider the two and higher loop contributions: the dominant loops, when resummed, give Eq.~(\ref{GabHighk}), again saying that all dominant loops have momenta of order $q\simeq \sigma_v^{-1}$ flowing in them; this is a consequence of the form of the vertices and the shape of the CDM spectrum. 

As far as the subdominant loops are concerned, there are two cases at the two-loop level: with one leg in the principal line (e.g. as the dotted loop in the top half of the $G_{ab}$ diagram in Fig.~\ref{PrincipalTrees}), and with no legs in the principal line (e.g. as the dotted loop in the bottom half of the $G_{ab}$ diagram in Fig.~\ref{PrincipalTrees}). In fact, there are two types of diagrams of the latter type, depending on whether the two legs of the loop land are on the same side of $\otimes$ (as in Fig.~\ref{PrincipalTrees}), or not. The first case corresponds to renormalizing the internal propagator, the second to a renormalization of the initial power spectrum. However, since both renormalizations must be evaluated at large-scales, i.e. $q \la \sigma_v^{-1}$, their effect on the final result must be subdominant. The subdominant loops with one leg in the principal line corresponds to renormalizing the vertex, and the level of agreement seen in Fig.~\ref{LargeKlimit} suggests that such vertex renormalizations are small. This corresponds to studying the three-point propagator in the limit where one incoming momentum is ``soft" (of order $q \la \sigma_v^{-1}$) while the other is large. While we are not able to resum  this case, we shall see that when the whole principal tree is in the high-$k$ limit, the behavior of the three-point propagator can be established.

\subsection{The Large-$k$ Limit of the Three-Point Propagator}
\label{Gamma2Res}

The aim of this section is to obtain the large-$k$ limit of the three-point propagator $\Gamma^{(2)}$ defined in Eq.~(\ref{GammaabcDef}). To this end we need to sum over the whole set of dominant loop diagrams,
\begin{equation}
\Gamma^{(2)}_{abc}=\sum_{n\ge 0} \Gamma^{(2)}_{abc,\ \rm n-loops},
\label{sumloops}
\end{equation}
that is, only consider those diagrams with loops directly attached to the principal tree, see Fig.~\ref{Gamma2Loops}. Note that to do the summation in Eq.~(\ref{sumloops}) one needs to take into account the symmetry factors of each contribution~\footnote{An alternative approach has been advocated in~\cite{2007JCAP...06..026M} based on a Renormalization Group Theory like approach. Such a method aims at transforming discrete sums we have to deal with into a simple first order differential equation, the solution of which can be easily obtained. The two approaches are based on the same basic properties of the loop terms. We comment more on this in the appendices.}. Let us start by considering the one-loop case in detail first, since it will allow an easy extrapolation when accounting for arbitrary number of loops in the large-$k$ limit. 

Diagrams are computed according to the rules given in section \ref{nonlineardynamics}, but let us be a bit more precise in the notation. In Fig. \ref{Gamma2Loops}, each of the contributing loops carries a wave vector $\vq$ such that $q\simeq \sigma_v^{-1} \ll k_{1}$ and similarly $q\ll k_{2}$. It is connected to two segments (not necessarily distinct) taken in the set of three - denoted $(1)$, $(2)$ and $(3)$ in Fig. \ref{Gamma2Loops} - each carrying respectively the wave vectors $\vk_{1}$, $\vk_{2}$ and $\vk_{3}$. In the calculation of those diagrams one has to sum over all indices, describing density and velocity divergence perturbations (we implicitly assume that such indices when repeated are summed over). As an example, we write down the corresponding integral when the loop is connected to branches $3$ and $1$,
\begin{widetext}
\begin{eqnarray}
\Gamma^{(2)}_{abc,\{13\}}(\vk_3,\vk_1,\vk_2;s)=\int \!\! P_0(q)\,\dd^3q \int_{s'}^s \!\!\! \dd s_1 \int_0^{s}\!\!\!  \dd s' \int_0^{s'} \!\!\! \dd s_2\,g_{ad}(s-s_1)\,\gamma_{def}(\vk_3,\vq,\vk_3-\vq)\,e^{s_1} u_e\,g_{fh}(s_1-s') \nonumber \\
\times \gamma_{hij}(\vk_3-\vq,\vk_1-\vq,\vk_2)\,g_{ik}(s'-s_2)\,\gamma_{klm}(\vk_1-\vq,-\vq,\vk_1)\,e^{s_2} u_l\,g_{mb}(s_2)\,g_{jc}(s') 
\label{examplegamma}
\end{eqnarray}
\end{widetext}
where $s_1,s_2$ are the interaction times where the loop legs attach to the principal tree, and $s'$ is the interaction time belonging to the principal tree. Furthermore, we assumed growing mode initial conditions, so the initial spectrum $P_{ab}$ reduces to $u_a u_b P_0$, see Eq.~(\ref{Spectra}), and $g_{ab}(s)u_b=e^s u_a$.

It is crucial to note that, as for the calculation of $G_{ab}$, in the high-$k$ limit the vertex matrix takes the form~\cite{2006PhRvD..73f3520C},

\begin{equation}
\gamma_{abc}(\vk_i,\vq,\vk_i-\vq) u_b\approx \frac{\vk_{i}.\vq}{2\, q^2}\ \delta_{ac},
\label{tgamma}
\end{equation}
when contracted with an incident growing mode (such as those coming from the initial conditions). The interaction then leaves the modes along the principal tree lines unchanged and carrying the wave vector $\vk_{i}$ (since $\vk_{i}+\vq\approx \vk_{i}$) all along. As we now show, applying this large-$k$ limit to Eq.~(\ref{examplegamma}) simplifies calculations greatly.

Let us be general now and  consider a loop that joins the segment $(i)$ and $(j)$. Again, we denote the interaction times for such a loop as $s_{1}$ and $s_{2}$. The loop  contribution will eventually be given by the integration of these times over the time range of the corresponding segments. The integral over the segment ($i$) can be defined through the respective characteristic functions $\mI_{i}$, with $\mI_{1}(t)=\mI_{2}(t)=\Theta(t)\Theta(s'-t)$ for interactions after the initial conditions but before the vertex of the principal tree (at time $s'$) and $\mI_{3}(t)=\Theta(t-s')\Theta(s-t)$  for interactions after $s'$ and before the time of interest $s$ ($s\geq s'\geq 0$).

\begin{widetext}
The one-loop contribution is then given by
\begin{eqnarray}
\Gamma^{(2)}_{abc,\{ij\}}&=&-s_{ij}\int_{0}^{s}\dd s' g_{ad}(s-s')\gamma_{def}(\vk_{1},\vk_{2},\vk_{3})
g_{eb}(s')g_{fc}(s')\int\dd
s_{1}\mI_{i}(s_{1})\int\dd s_{2}\mI_{j}(s_{2})\nonumber\\
&&\times \int\dd^2\Omega_{q}q^2\dd
q\,P_{0}(q)\ \frac{\vk_{i}.\vq}{2\,q^2}\,\frac{\vk_{j}.\vq}{2\,q^2}\ e^{s_{1}+s_{2}},
\label{G2Loopij1}
\end{eqnarray}
where $s_{ij}$ is the symmetry factor associated to this contribution. It is 4 if $i\ne j$ (e.g. 2 for each incoming line), and it is only 2 for $i=j$ because the time ordering of $s_{1}$ and $s_{2}$ is ignored in Eq.~(\ref{G2Loopij1})~\footnote{We are able to compute time integrations ignoring time-ordering and correcting for this afterwards because the integrand (${\rm e}^{s_1+s_2}$) is symmetric under exchange of $s_1,s_2$. That is, for such integrands we can take advantage of $\int_{s'}^s ds_2 \int_{s'}^{s_2}  ds_1 = {1\over 2} \int_{s'}^s ds_2 \int_{s'}^s ds_1$. This symmetry arises as a consequence of the simplification of the vertex function in the high-$k$ limit, plus the linear propagator obeying $g_{ab}(s-s')g_{bc}(s')=g_{ac}(s)$. We will use this useful property in the summation over loops shortly, as it is more convenient to deal with an arbitrary number of integrals with the same limits.}. The integral over the angles $\Omega_{q}$ is straightforward and leads to,
\begin{eqnarray}
\Gamma^{(2)}_{abc,\{ij\}}&=&-\frac{s_{ij}}{4}\int_{0}^{s}\dd s' g_{ad}(s-s')\gamma_{def}(\vk_{1},\vk_{2},\vk_{3})
g_{eb}(s')g_{fc}(s')\int\dd s_{1}\mI_{i}(s_{1})\int\dd
s_{2}\mI_{j}(s_{2})\ \vk_{i}.\vk_{j}\ \sigma_{v}^2\ e^{s_{1}+s_{2}},
\label{G2Loopij2}
\end{eqnarray}
where $\sigma_{v}^2$ is defined in Eq.~(\ref{sigmavdef}). Note that, if we could ignore the last two time integrals, the answer would be proportional to the tree-level amplitude (given by the first integral). The time integration over $s_{1}$ and $s_{2}$ can be done easily for each loop diagram characterized  by $\{ij\}=\{11\},\{22\},\{33\},\{12\},\{13\},\{23\}$.  When we collect all the one-loop contributions we get,
\begin{eqnarray}
\Gamma^{(2)}_{abc,\ \rm{1loop}}&=&\sum_{ij}\Gamma^{(2)}_{abc,\{ij\}} \nonumber \\
&=&-\frac{1}{2}\int_{0}^{s}\dd s' g_{ad}(s-s')\gamma_{def}(\vk_{1},\vk_{2},\vk_{3})
g_{eb}(s')g_{fc}(s')\ \sigma_{v}^2\nonumber\\
&&\times
\left[\left(e^{s'}-1\right)^2\left(k_{1}^2+2\vk_{1}.\vk_{2}+k_{2}^2\right)
+2\left(e^s-e^{s'}\right)\left(e^{s'}-1\right)\left(\vk_{1}.\vk_{3}+\vk_{2}.\vk_{3}\right)+
\left(e^s-e^{s'}\right)^2k_{3}^2\right],
\label{G2intermediate}
\end{eqnarray}
\end{widetext}
which when using that $\vk_3=\vk_{1}+\vk_{2}$ makes the square bracket independent of $s'$ and leads to the remarkable result:
\begin{equation}
\Gamma^{(2)}_{abc,\ \rm{1loop}}=-\frac{k_{3}^2\,\sigma_{v}^2}{2}\ \Gamma^{(2)}_{abc,\ \rm{tree}}\
\left(e^{s}-1\right)^2.
\label{G21Loop}
\end{equation}
We are now ready to consider diagrams with an arbitrary number of loops. As we shall see it amounts to exponentiate the one-loop result above. There are at least two ways to obtain such a result. The most direct method, which we follow here, consists in computing simultaneously an arbitrary number of loops. The cumbersome part in this case is to find out the combinatorial weights for each diagram. Here we follow the one-loop strategy discussed above and ignore time ordering at first and correct for the overcounting afterwards. In Appendix~\ref{TOrd} we provide an alternative derivation of this method that takes into account time ordering explicitly. A second method based on a different resummation scheme is presented in Appendix~\ref{ResFS}. 

The lesson we must recall from the one-loop calculation is the following. Every loop must be directly connected to two ``principal'' segments $(i)$ and $(j)$ of the principal tree ($i$ and $j$ go from 1 to 3, see Fig.~\ref{Gamma2Loops}). Time integration then leads to $(e^{s'}-1)$ if $i=1$ or $2$, and $(e^s-e^{s'})$ if $i=3$, and similarly for $j$. Momentum integration leads to $-\sigma_v^2\,\vk_i \cdot \vk_j/4$, and in addition there is an overall symmetry factor $s_{ij}$ (with $s_{ij}=4$ if $j\neq i$ and 2 otherwise).

\begin{figure*}
\centerline{\epsfig {figure=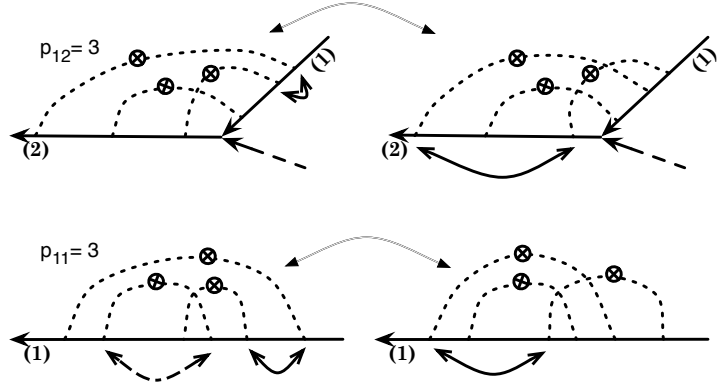,width=10cm}}
\caption{This figure illustrates the effect of the time ordering exchanges (thick double arrow lines).\\
Through such exchanges, the complete set of diagrams that correspond to a given $\{p_{ij}\}$ can be explored. Successive time exchanges can however lead to identical diagrams (e.g. left and right diagrams in each panel). The unordered time integration then leads to a multiplicity factor for each diagram. For the top panel (corresponding to $p_{12}=3$), the same diagram is obtained each time two loop lines are exchanged. There are $p_{12}!$ of such possible exchanges. For the bottom panel (corresponding to $p_{11}=3$), the same diagram can be obtained either by the exchange of loop lines, or by the exchange of the end and final time of each loop of them (double arrow dashed line). There are thus $p_{11}!\ 2^{p_{11}}$ of such possible exchanges.
} 
\label{Gamma2Sym}
\end{figure*}

The loop resummation in Eq.~(\ref{sumloops}) relies on a proper counting of each of the diagrams that contribute in the large-$k$ limit. In turn, each of these diagrams can be defined by a set of parameters, $p_{ij}$, where each counts the number of lines connecting the segments $(i)$ and $(j)$ ($i\le j$). For instance, the diagram in Fig.~\ref{Gamma2Loops} corresponds to $p_{11}=1$, $p_{22}=0$, $p_{12}=1$, $p_{13}=1$, $p_{23}=0$ and $p_{33}=1$. Each loop line connects to the principal tree at time $s_{i}$. Dropping the time ordering in $s_{i}$ (within each of the principal lines), the contribution from such an ensemble of loops is given by,
\begin{widetext}
\begin{eqnarray}
\Gamma^{(2)}_{abc,\{p_{ij}\}}&=&\frac{s_{\left\{p_{ij}\right\}}}{\mM_{\left\{p_{ij}\right\}}}
\left(-\frac{\sigma_{v}^2}{4}\right)^{\sum_{i\le j}p_{ij}}\prod_{i} k_{i}^{2p_{ii}}\prod_{i<j}
\left(\vk_{i}.\vk_{j}\right)^{p_{ij}}
\int_{0}^{s}\dd s' g_{ad}(s-s')\gamma_{def}(\vk_{1},\vk_{2},\vk_{3})
g_{eb}(s')g_{fc}(s')\nonumber\\
&&\times \left(e^{s'}-1\right)^{2p_{11}+2p_{22}+2p_{12}+p_{13}+p_{23}}
\left(e^s-e^{s'}\right)^{2p_{33}+p_{13}+p_{23}}.
\label{G2Loops1}
\end{eqnarray}
\end{widetext}
In this expression the coefficient $s_{\left\{p_{ij}\right\}}$ is the symmetry factor associated with
each set of $p_{ij}$. For a given ordered topology, each loop introduces a factor 4, so that
\begin{equation}
s_{\left\{p_{ij}\right\}}=2^{2\sum_{i\le j}p_{ij}}.
\label{sijexp}
\end{equation}
The coefficient ${\mM_{\left\{p_{ij}\right\}}}$ on the other hand represents the number of times each given diagram has been computed because the time ordering has been dropped. Changing the time
order (within each principal line) allows to automatically scan the different topologies (diagrams sharing the same set of $p_{ij}$) but computing them several times. The symmetry factor  $p_{ij}$ in ${\mM_{\left\{p_{ij}\right\}}}$ is precisely here to correct for this multiplicity factor. It is relatively easy to see that its dependence on each $p_{ij}$ factorizes, so that
\begin{equation}
{\mM_{\left\{p_{ij}\right\}}}=\prod_{i\le j}\mM(p_{ij})
\label{Cij}
\end{equation}
where $\mM(p_{ii})=2^{p_{ii}}p_{ii}!$, and $\mM(p_{ij})=p_{ij}!$ if $i\ne j$
(there is an extra factor of 2 when $i=j$ to compensate for the absence of time ordering. See Fig.
\ref{Gamma2Sym} for illustration of these symmetry factor countings). The summation over $p_{ij}$ is then straightforward and leads to exponentiations,
\begin{widetext}
\begin{eqnarray}
\Gamma^{(2)}_{abc}=\sum_{\left\{p_{ij}\right\}}\Gamma^{(2)}_{abc,\{p_{ij}\}} &=&
\int_{0}^{s}\dd s' g_{ad}(s-s')\gamma_{dbc}(\vk_{1},\vk_{2},\vk_{3})
g_{db}(s')g_{dc}(s')\nonumber\\
&&\hspace{-2cm}\times\exp\left\{-\frac{\sigma_{v}^2}{2}
\left[\left(e^{s'}-1\right)^2\left(k_{1}^2+2\vk_{1}.\vk_{2}+k_{2}^2\right)
+2\left(e^s-s^{s'}\right)\left(e^{s'}-1\right)\left(\vk_{1}.\vk_{3}+\vk_{2}.\vk_{3}\right)+
\left(e^s-e^{s'}\right)^2k_{3}^2\right]\right\}\nonumber\\
&=&\exp\left(-\frac{\sigma_{v}^2k_{3}^2}{2}\left(e^s-1\right)^2\right)\
\Gamma^{(2)}_{abc,{\rm tree}},
\label{G2Loops2}
\end{eqnarray}
\end{widetext}
with $\Gamma^{(2)}_{abc,\rm tree}$ defined in Eq.~(\ref{Gamma2Tree}). This is a truly remarkable result. It shows that the whole effect of loop summation  is encoded in the value of $k_{3}$ in exactly the same way as for the two-point propagator $G_{ab}$. 

\begin{figure}
\centerline{\epsfig {figure=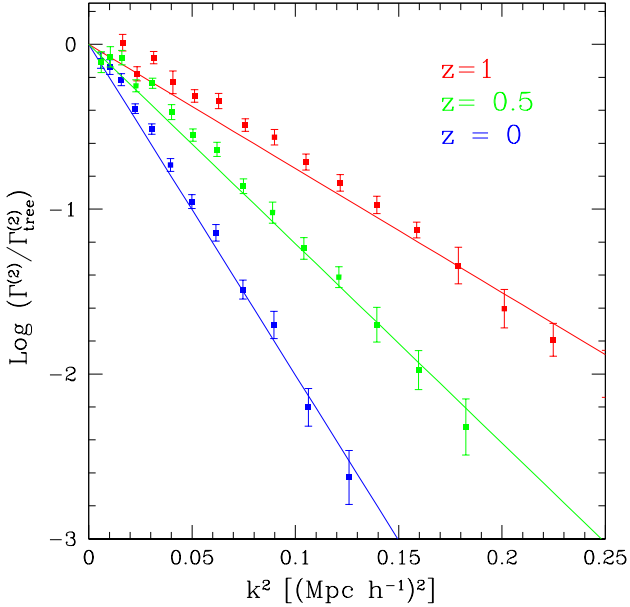,width=8.5cm}}
\caption{The large-$k$ limit of the three-point density propagator $\Gamma^{(2)}_1\equiv \Gamma^{(2)}_{1bc} u_b u_c$, the only density contraction that can be measured for growing mode initial conditions,   $u_b=(1,1)$. The symbols in the figure correspond to  equilateral configurations at redshifts $z=1, 0.5, 0$ (from top to bottom). We have normalized these measurements to its low-$k$ limit $\Gamma^{(2)}_{1,\rm tree}$ given by Eq.~(\ref{Gamma1tree}). The figure clearly shows that the measured propagator closely follows the large-$k$ limit given by Eq.~(\ref{G2Loops2}) represented by solid lines, once $\Gamma^{(2)}_1$ decays by $\approx {\rm e}^{-1}$ from its tree-level value.} 
\label{Gamma2LargeKlimit}
\end{figure}

We now compare this result to measurements in numerical simulations, which will be described in detail in section~\ref{numsimu}. As done for the two-point propagator (see Fig.~\ref{LargeKlimit}), we test for the Gaussian decay in the high-$k$ limit by plotting in Fig.~\ref{Gamma2LargeKlimit} equilateral configurations $\log \Gamma^{(2)}_1(k,k,k)$ vs $k^2$, for which Eq.~(\ref{G2Loops2}) predicts a straight line with known slope. We do so for three different redshifts $z=0,0.5,1$, finding very good  agreement in all cases with the predictions of Eq.~(\ref{G2Loops2}), shown in solid lines.  This validates our resummation scheme. 
  
Equation~(\ref{G2Loops2}) and its generalization to other multi-point propagators have important implications for the power spectrum and higher-order statistics, that we discuss in section~\ref{recon}. We note also that a second, faster,  method to perform the loop resummation is discussed in appendix~\ref{ResFS}. We now consider the extension of these results to arbitrary multi-point propagators. 

\subsection{The Large-$k$ Limit for Higher-Order Multi-Point Propagators}

The structure we found for the three-point propagator $\Gamma^{(2)}$ is appealing enough to consider its full generalization to propagators of arbitrary number of points. The crucial property is the extension of the one-loop relation given in Eq.~(\ref{G21Loop}). For higher than three-point propagators, the tree-order is given by the sum of several diagrams. When loop corrections  are taken into account, the {\em principal tree} of each of these diagrams is naturally the corresponding tree-order diagram. For instance, there are two of such diagrams for $\Gamma^{(4)}$ shown in Fig. \ref{DiagramPsiExpansion}.

Let us define $\Gamma^{(p)}_{(q),{\rm tree}}$, the tree-order contribution to $\Gamma^{(p)}$ from diagram of type $(q)$. The value of such a diagram is obtained after the integration over the $s_{1}\dots s_{p-1}$ time variables corresponding to the interaction points is done. Generically it can be written as,
\begin{eqnarray}
\Gamma^{(p)}_{(q),{\rm tree}}=\int_{0}^{s}\dd s_{1}\dots \int_{0}^{s}\dd
s_{p-1}\mT^{(q)}(\vk_{1},\dots,\vk_{p-1}, \nonumber \\ ,s_{1},\dots, s_{p-1})
\label{Gammapqtree}
\end{eqnarray}
where $\mT^{(q)}(\vk_{1},\dots,\vk_{p-1},s_{1},\dots, s_{p-1})$ denotes the value of the diagram $(q)$ before the time integrations are performed. $\mT$ depends on the vertices and the linear propagators, the arguments of which impose time-ordering through $g_{ab}(\eta)=0$ for $\eta<0$. 

\begin{figure*}
\centerline{\epsfig {figure=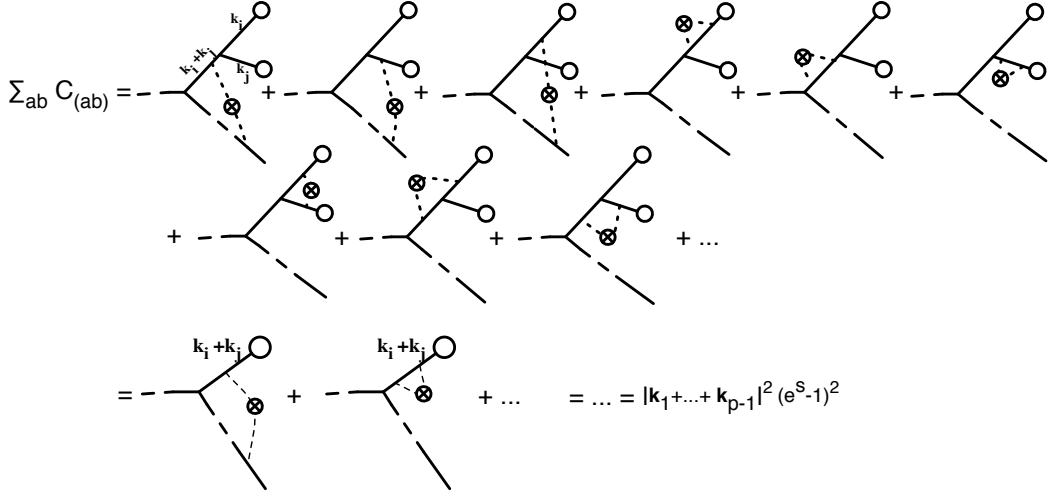,width=14cm}}
\caption{Sum of the one-loop contributions described by $\mC_{(ij)}$, defined in Eq.~(\ref{Gammap1loop}). The contributions involving the wave vectors $\vk_{i}$, $\vk_{j}$ and their sum is singled out. The first  three diagrams correspond to the one-loop contribution involving an extra wave vector, $\vk'$ (dashed line). By the high-$k$ summation rules described before Eq.~(\ref{G2Loops1}), they sum to $\vk'\cdot\left[(\vk_{i}+\vk_{j})(e^{s_{2}}-e^{s_{1}})+ \vk_{i}(e^{s_{1}}-1)+\vk_{j}(e^{s_{1}}-1)\right](e^{s_{2}}-1)=\vk'\cdot(\vk_{i}+\vk_{j})(e^{s_{2}}-1)^2$. The next three and last three diagrams correspond to ``internal" loops (i.e. loops attached to the principal lines with momenta $\vk_{i}$, $\vk_{j}$ and their sum). As shown for the computation of $\Gamma^{(2)}$, these six diagrams sum to  $\vert\vk_{i}+\vk_{j}\vert^2 (e^{s_{2}}-1)^2$. By the  summation rules, these two results can be expressed by two contributions to $\mC_{(ij)}$ with one less branch, as shown by the two diagrams in the last line. The same procedure can be applied recursively until the expression of $\mC_{(ij)}$ reduces to that coming from a single branch, leading to the final expression.}
\label{GammapReduction}
\end{figure*}

The one-loop diagrams built out of the $(q)-$tree diagram involve the same core function $\mT^{(q)}(\vk_{1},\dots,\vk_{p-1},s_{1},\dots, s_{p-1})$. A one-loop diagram joining the segments $(i)$ and $(j)$ is then given by
\begin{widetext}
\begin{eqnarray}
\Gamma^{(p)}_{(q),\{ij\}}&=&\frac{-\sigma_{v}^2}{2}\int_{0}^{s}\dd s_{1}\dots \int_{0}^{s} \dd
s_{p-1}\ \mT^{(q)}(\vk_{1},\dots,\vk_{p-1},s_{1},\dots, s_{p-1})\ \mC_{\{ij\}}(\vk_{1},\dots,\vk_{p-1},s_{1},\dots,
s_{p-1}) \\
\mC_{\{ij\}}&\equiv &\frac{s_{ij}}{2}
(\vk_{i}\cdot\vk_{j})\left(e^{s_{i,2}}-e^{s_{i,1}}\right)\left(e^{s_{j,2}}-e^{s_{j,1}}\right)
\label{Gammap1loop}
\end{eqnarray}
\end{widetext}
where $s_{i,1}$ and $s_{i,2}$ are respectively the starting and ending time of segment $(i)$ of the principal tree which carries momentum $\vk_i$. Thus the times $s_{k,l}$ belong to the set $\{s_1,\ldots ,s_{p-1}\}$.  This is already the result of integrating over the additional two time variables that correspond to the two extra interactions due to the loop (see the step in going from Eq.~(\ref{G2Loopij2}) to Eq.~(\ref{G2intermediate}), for example). 

From these results it can then be shown that the sum over $(i,j)$ of $\mC_{\{ij\}}$ is $\vert \vk_{1}+\dots +\vk_{p-1}\vert^2 (e^s-1)^2$. The proof can be obtained recursively as a simple successive application of the result found for $\Gamma^{(2)}$. This mechanism is illustrated in Fig. \ref{GammapReduction}. It leads to ($\vk_p=\vk_{1}+\dots +\vk_{p-1}$),
\begin{equation}
\Gamma^{(p)}_{(q),{\rm 1loop}}=-\frac{k_{p}^2\sigma_{v}^2}{2}(e^s-1)^2
\Gamma^{(p)}_{(q),{\rm tree}}.
\label{Gammapq1loop}
\end{equation}
As shown in appendix~\ref{ResFS}, the scheme for resummation over all loop terms can be applied once again, leading to the generalization  of Eq.~(\ref{G2Loops2}),
\begin{equation}
\Gamma^{(p)}=\exp\left(-\frac{k_{p}^2\sigma_{v}^2}{2}(e^s-1)^2\right)\ 
\Gamma^{(p)}_{{\rm tree}}.
\label{Gammapq1loops}
\end{equation}

\section{One-Loop Results for the Three-Point Propagator}
\label{oneloopcase}

Having derived expressions for large scales, the tree-level expression in Eq.~(\ref{Gamma2Tree}), and small scales (high-$k$ limit), we now discuss what happens in between. This can be achieved by computing the next to leading order correction in an series expansion in terms of initial fields, i.e. $p=1$ in Eq.~(\ref{GammaabcExp2}). Diagrammatically it corresponds to calculating the one-loop diagrams from Fig.~\ref{Gamma2Loops}, according to the rules given in section~\ref{nonlineardynamics}. There are six types of such diagrams, depending on where each of the loop lines is attached to in the principal tree. Depending on this, they can be labelled as $\{33\},\{11\},\{22\},\{12\},\{13\}$, and $\{23\}$. Each of these cases is depicted in Fig.~\ref{Gamma2Loops}. 

In Eq.~(\ref{examplegamma}) we wrote down explicitly the one-loop integral expression for $\Gamma^{(2)}_{abc,\{13\}}$. A similar procedure can be followed to obtain expressions for the rest of the one-loop contributions. The difficulty in their actual computation resides in the integration over loop momenta. The angular integrations are independent of $P_{0}(k)$ and can be done analytically. Such integrations involve two sets of functions, one built out of
\begin{equation}
\int\dd^{2}\Omega_{q}\,\frac{1}{\vert \vk_{1}+\vq\vert^2}=
\frac{\pi  \log}{k_1 q} \left(\frac{(k_1+q)^2}{(k_1-q)^2}\right)
\label{Ha00}
\end{equation}
and the other built out of,
\begin{eqnarray}
\int\dd^{2}\Omega_{q}\,\frac{1}{\vert \vk_{1}+\vq\vert^2\,\vert \vk_{2}-\vq\vert^2} = \ \ \ \ \ \ \  \nonumber \\ 
\frac{\pi \log\left(\mW_{+}/\mW_{-}\right)}{k_3\, q\,
\sqrt{k_1^2 k_2^2 + q^4 + q^2(k_3^2-k_1^2-k2^2)}}   
\label{Hd00}
\end{eqnarray}
with,
\begin{eqnarray}
\mW_{\pm}\equiv \pm 2(k_1^2-q^2)(k_2^2-q^2) \pm 4 k_3^2 q^2+ 4 k_3 q \nonumber \\ \times \sqrt{k_1^2 k_2^2 + q^4 + q^2(k_3^2-k_1^2-k2^2)} 
\end{eqnarray}
and $k_3=|\vk_1+\vk_2|$. The explicit expression of the final result for arbitrary dependence on configuration ($k_1,k_2,k_3$) is too long to reproduce here. However, we have checked that $\Gamma^{(2)}_{abc,\rm{1loop}}(\vk_{1},\vk_{2},q)$ follows the expected large-$k$ limit.  In the low-$k$ limit, it is interesting to note that after the angular integration $\Gamma_{abc,\rm{1loop}}(\vk_1,\vk_{2},q)$ behaves as $1/q^2$. 
\begin{widetext}
The most growing contribution (i.e. largest power in growth factor) in this limit is,
\begin{eqnarray}
\Gamma^{(2)}_{1,{\rm 1loop}}(\vk_1,\vk_{2},s)&\sim& - \left[
\frac{4901(k_1^2+k_2^2)}{18865}+\frac{32879 k_1^4 + 231478 k_1^2 k_2^2 + 32879 k_2^4}{226380 k_1^2 k_2^2} \,(\vk_1\cdot\vk_2) \nonumber \right. \\ &&  \left. +\frac{9552(k_1^2+k_2^2)}{18865 \, k_1^2 k_2^2} \,(\vk_1\cdot\vk_2)^2 +\frac{12409}{56595\,k_1^2 k_2^2}\,(\vk_1\cdot\vk_2)^3 \right] \sigma_v^2 a^4(\tau),
\label{G2lowk}
\end{eqnarray}
with $\sigma_v^2$ given in Eq.~(\ref{sigmavdef}) and $a(\tau)$ the linear growth factor. Note that its angular dependence can be different from that of the tree order expression, depending on configuration.

For the complete result (but still most growing) valid at all scales we restrict to particular triangle configurations. For {\em equilateral configurations} ($k_{1}=k_{2}=k_{3}=k$),
\begin{eqnarray}
\Gamma_{1,{\rm 1loop}}^{(2)~{\rm equi.}}&=& \int P_0(q) \ \dd^3 q \
\frac{1}{2483712
   k^5 q^5 \sqrt{k^4-q^2 k^2+q^4}}   \left\{4k
  \left[6 \left(32 k^2+45 q^2\right)
   \log \left({\mW_{+}^{{\rm equi}}}/{\mW_{-}^{{\rm equi}}}\right)
   \left(k^3-k q^2\right)^3\right.\right.
   \nonumber\\
   &&\left. \left. +q \sqrt{k^4-q^2 k^2+q^4}
   \left(5754 k^8-33521 q^2 k^6+41247 q^4 k^4-27039 q^6
   k^2+5175 q^8\right)\right] \right.
   \nonumber\\
   &&\left.   -3 \left(k^2-q^2\right)^3
   \left(2174 k^4+4413 q^2 k^2-1725 q^4\right)
   \sqrt{k^4-q^2 k^2+q^4} \log
   \left(\frac{(k+q)^2}{(k-q)^2}\right)\right\},
\label{oneloopequilateral}
\end{eqnarray}
where $\mW_{\pm}^{{\rm equi}}$ correspond to the value of $\mW_{\pm}$ for an equilateral
configuration. For {\em colinear configurations} ($k_{3}=2k_{1}=2k_{2}=k$),
\def\LWCols{\log(\mW_{+}^{(1)}/\mW_{-}^{(1)})}
\def\LWColl{\log(\mW_{+}^{(2)}/\mW_{-}^{(2)})}
\begin{eqnarray}
\Gamma_{1,{\rm 1loop}}^{(2) ~{\rm colin.}}&=& \int P_0(q) \ \dd^3 q \ \left\{
\frac{57171840 k^8 - 69330880 q^2 k^6 + 49644768 q^4 k^4 - 13102080 q^6 k^2 + 
      2349765 q^8}{3179151360 k^4 q^4} + \right. \nonumber \\
&&      
      \frac{3 \LWCols \left(q^2 - 4 k^2\right)^4}{275968 k 
              q^5 \sqrt{16 k^4 - 7 q^2 k^2 + q^4}}+
\frac{{\LWColl} \left(q^2 - 4 k^2\right)^2 \left(2 k^4 + 7 
              q^2 k^2 - 9 q^4\right)}{29568 k 
        q^5 \sqrt{4 k^4 - q^2 k^2 + q^4}}-\nonumber\\
        &&       
    \frac{\left(1146 k^{10} + 3259 q^2 k^8 - 7539 q^4 
            k^6 + 4191 q^6 k^4 - 1363 q^8 
            k^2 + 306 q^{10}\right) \log \left(\frac{(k + q)^2}{(k - q)^2}\right)}{1892352 k^5
q^5}-
\nonumber\\ 
&&  \left.
\frac{\left(q^2 - 4 k^2\right)^2 \left(976064 k^6 - 253104 q^2 k^4 - 
        178124 q^4 k^2 + 19563 q^6\right) \log \left(\frac{(2 k + q)^2}{(q - 
            2 k)^2}\right)}{1695547392 k^5 q^5} \right\}  
\label{oneloopcolinear}
\end{eqnarray}
where $\mW_{\pm}^{(1)}=\mW_{\pm}(2k,2k,k)$ and $\mW_{\pm}^{(2)}=\mW_{\pm}(2k,k,2k)$. 

\end{widetext}

Equations (\ref{oneloopequilateral}) and (\ref{oneloopcolinear}), fully specify the most growing contributions to the {\em density} propagator, their time dependence being an overall scaling by $a^4(\tau)$.

\begin{figure}
\centerline{\epsfig {figure=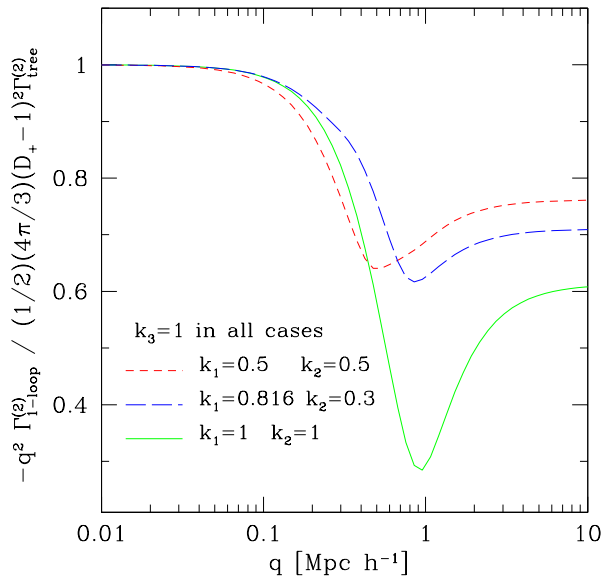,width=8cm}}
\caption{Momentum dependence of the three-point density propagator integrand $\Gamma^{(2)}_{1,{\rm 1loop}}(\vk_1,\vk_{2},q)$ for different configurations. The solid line denotes equilateral triangles (with $k_{1}=k_{2}=k_{3}=1$); the short dashed line colinear triangles (with $k_{1}=k_{2}=0.5,k_{3}=1$); and the long dashed line triangles with $k_{1}=0.816,k_{2}=0.3,k_{3}=1$ (in all cases units are in $h$/Mpc). 
The y-axis has been normalized to the large-$k$ limit expression given in Eq.~(\ref{G21Loop}). Notice how this limit is reached for small values of $q$, as expected.}
\label{Gamma21loop-qdep}
\end{figure}

As examples we show in Fig.~\ref{Gamma21loop-qdep} the dependence of $\Gamma^{(2)}_{1,{\rm 1loop}}(\vk_1,\vk_{2},q)$ on $q$ for particular triangular configurations, with $\Gamma^{(2)}_{1,{\rm 1loop}}(\vk_1,\vk_{2},q)$ defined such that $\Gamma^{(2)}_{1,1-\rm{loop}}(\vk_1,\vk_{2})=\int \Gamma^{(2)}_{1,{\rm 1loop}}(\vk_1,\vk_{2},q) P_0(q) q^2 \dd q $. For small values of $q$, corresponding to the large-$k$ limit, we recover the tree-order result.  For large values of $q$, the integrand behaves similarly, but with a different amplitude.

\section{The Three-Point Propagator in Numerical Simulations}
\label{numsimu}

We now show in more detail comparison of our results against measurements in numerical simulations for the three-point propagator. 
As mentioned before, the cross-correlation property, Eq.~(\ref{ThreeMomentExp}), allows us to easily measure the three-point propagator in numerical simulations. In fact, this can be straightforwardly extended to other multi-point propagators using Eq.~(\ref{npointexpression}). 

In practice initial conditions in simulations are generally set in the linear growing mode in which case Eq.~(\ref{ThreeMomentExp}) becomes,
\begin{eqnarray}
\mg\Psi_{a}(\vk)\delta_{0}(-\vk_{1})\delta_{0}(-\vk_{2})\md&=&2\,\Dirac(\vk-\vk_{1}-\vk_{2})
\nonumber\\
&&\hspace{-2cm}\times\ \Gamma^{(2)}_a(\vk_{1},\vk_{2})P_{0}(k_{1})P_{0}(k_{2}),
\label{GammaaExp3}
\end{eqnarray}
where we used the shorthand notation $\Gamma^{(2)}_a=\Gamma^{(2)}_{abc}u_b u_c$ with $u_a=(1,1)$ and $\phi_{a}(\vk)=u_{a}\delta_{0}(\vk)$, and the initial power spectrum obeys $P_0(k) \delta_{\rm D}\, (\vk+\vk')=\mg \delta_0(\vk) \delta_0(\vk') \md$).

Therefore, $\Gamma^{(2)}_a$ can be measured by implementing the following algorithm based on Eq.~(\ref{GammaaExp3}),
\begin{eqnarray}
\Gamma^{(2)}_a(k_1,k_2,k_3)=\frac{1}{2 P_0(k_1)P_0(k_2)}\frac{1}{N}\sum_{\vk_i \vk_j} \sum_{\vk_l}\Psi_a(\vk_l,s) \nonumber\\ 
\times\delta_0(-\vk_i)\delta_0(-\vk_i), \ \ \ \ \ 
\label{algorithm}
\end{eqnarray}
where the sum runs over Fourier modes $\vk_i$ in the $|\vk_1|$ bin, $\vk_j$ in the  $|\vk_2|$ bin and
$\vk_l$ in a bin $|\vk_3|$ such that $|\vk_1-\vk_2| \le |\vk_3| \le \vk_1+ \vk_2$, and $N$ is the number of terms in the triple sum.   

In writing Eq.~(\ref{algorithm}) we have assumed that the three-point correlator  $\Gamma^{(2)}_a$ does not depend upon the orientation of the triangle  but only on the magnitude of its sides, as expected from statistical isotropy. Additionally one must be aware that the estimator in Eq.~(\ref{algorithm}) is only strictly valid when the initial field $\delta_0$ is Gaussian (otherwise, one must go back to the definition~Eq.~(\ref{GammaabcDef}) and implement the functional derivative, see~\cite{2006PhRvD..73f3520C} for  such details in the case of the two-point propagator).

In this paper we focus on the statistics of the density field, and thus we only present measurements of $\Gamma^{(2)}_1$. Obtaining $\Gamma^{(2)}_2$ requires the measurement of the local velocity divergence,  while measuring other components of $\Gamma^{(2)}_{abc}$ requires numerical simulations with a mixture of growing and decaying modes in the initial conditions.

To test our theoretical results from sections \ref{largekGammap} and \ref{oneloopcase}, we use Eq.~(\ref{algorithm} to measure the three-point propagator in  N-body simulations using a set of $50$ realizations, with each realization containing   $N_{par}=640^3$ particles within a cubic volume of side $L_{box}=1280 h^{-1}{\rm Mpc}$. The total comoving volume of our simulations, approximately $105~(h^{-1} {\rm Gpc})^3$, is large enough to minimize cosmic variance. Cosmological parameters were chosen as $\Omega_m=0.27$, $\Omega_{\Lambda}=0.73$,  $\Omega_b=0.046$ and $h=0.72$. The corresponding initial power spectrum had scalar spectral index $n_s=1$ and was normalized to yield $\sigma_8=0.9$ when linearly evolved to $z=0$. The simulations were run using  Gadget2~\cite{2005MNRAS.364.1105S} with initial conditions set at $z_i=49$ using 2nd order Lagrangian Perturbation Theory (2LPT)~\cite{1998MNRAS.299.1097S,2006MNRAS.373..369C}. The output times were at $z=1,0.5,0$. At each output time we measured the two and three-point propagators by cross-correlating the density field at the desired time with the initial density field at $z_i=49$ (as described above and in \cite{2006PhRvD..73f3520C}). 

\begin{figure*}
\begin{center}
\includegraphics[width=0.4\textwidth]{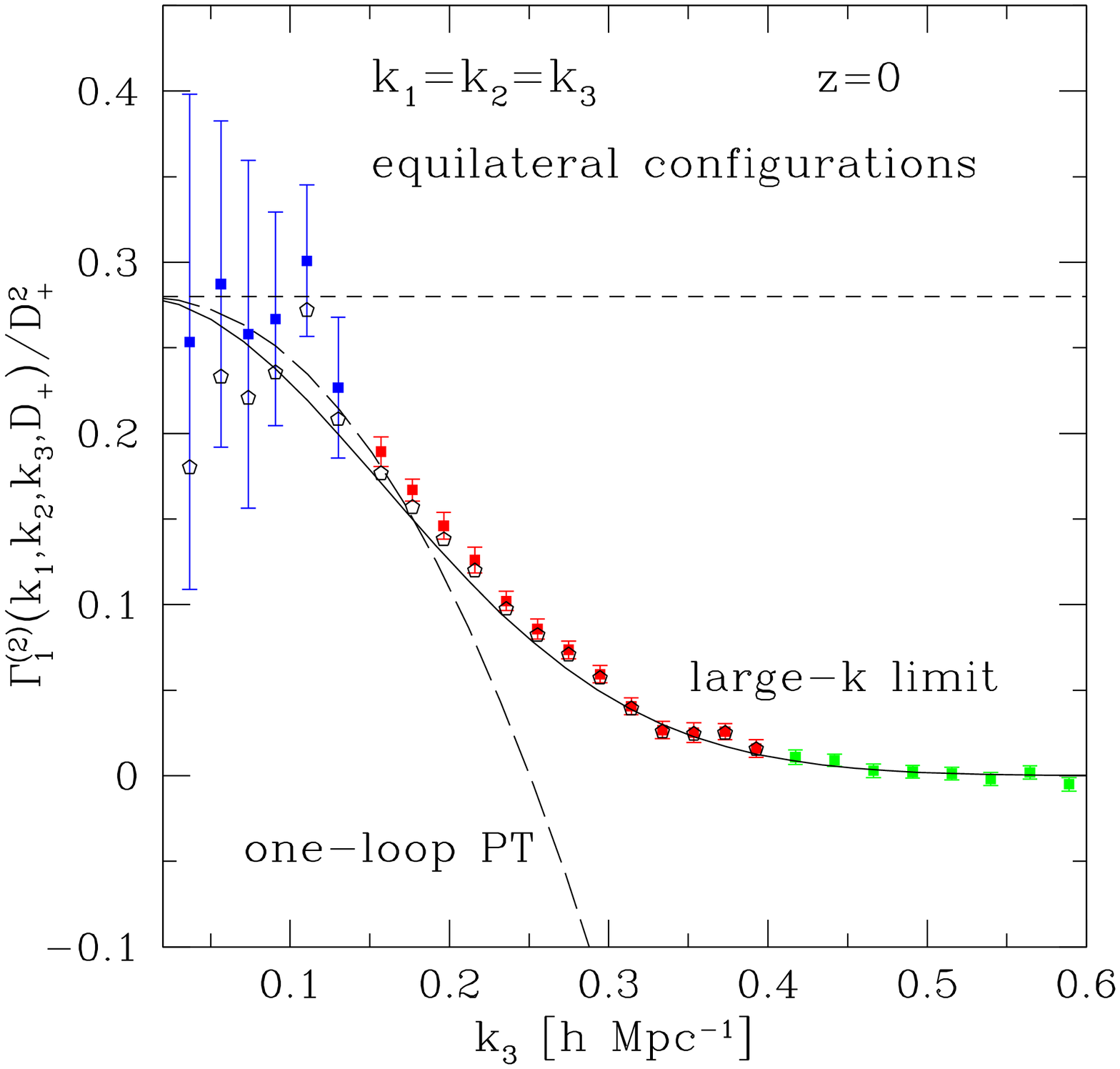}
\includegraphics[width=0.4\textwidth]{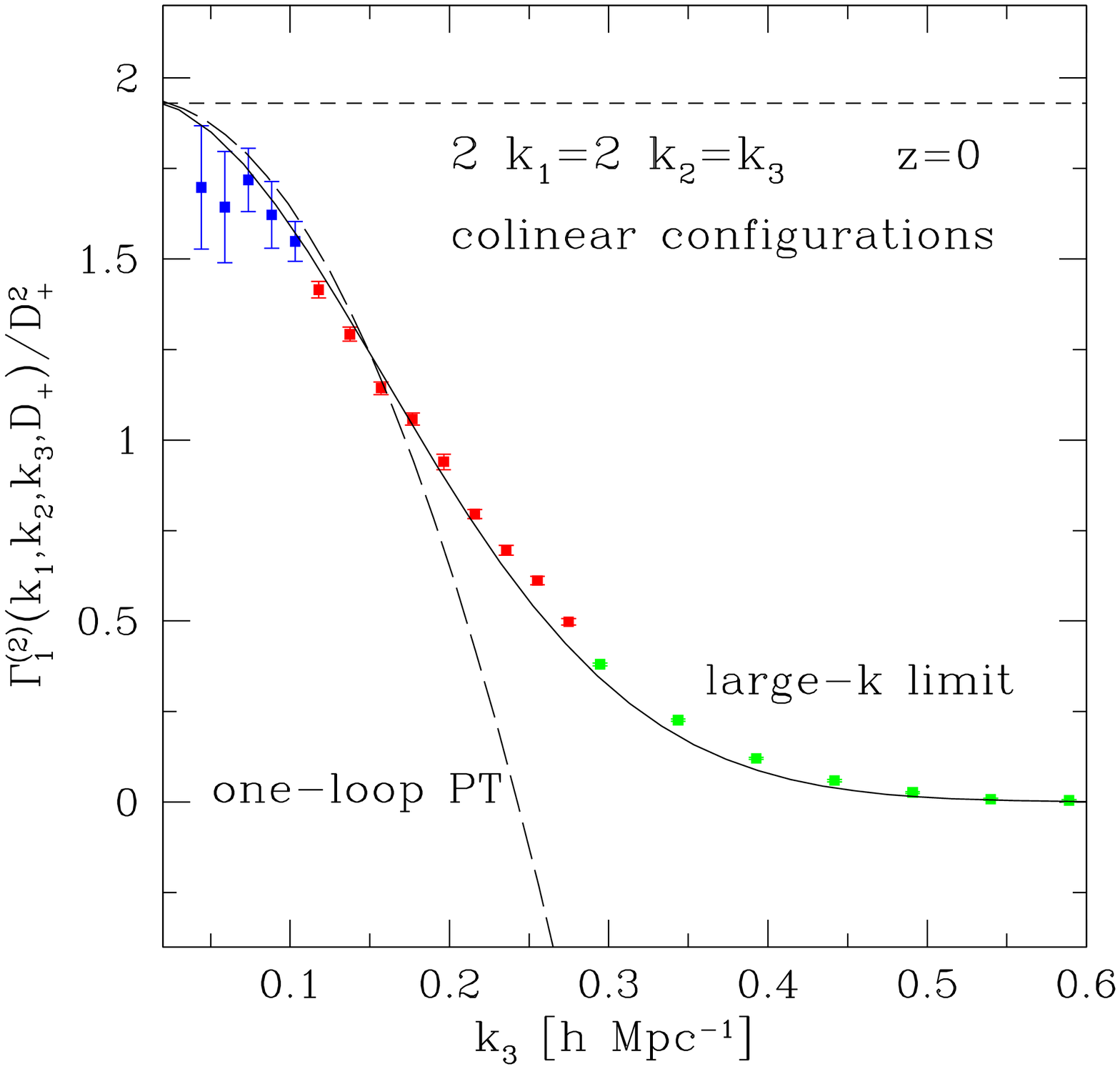} \\
\includegraphics[width=0.4\textwidth]{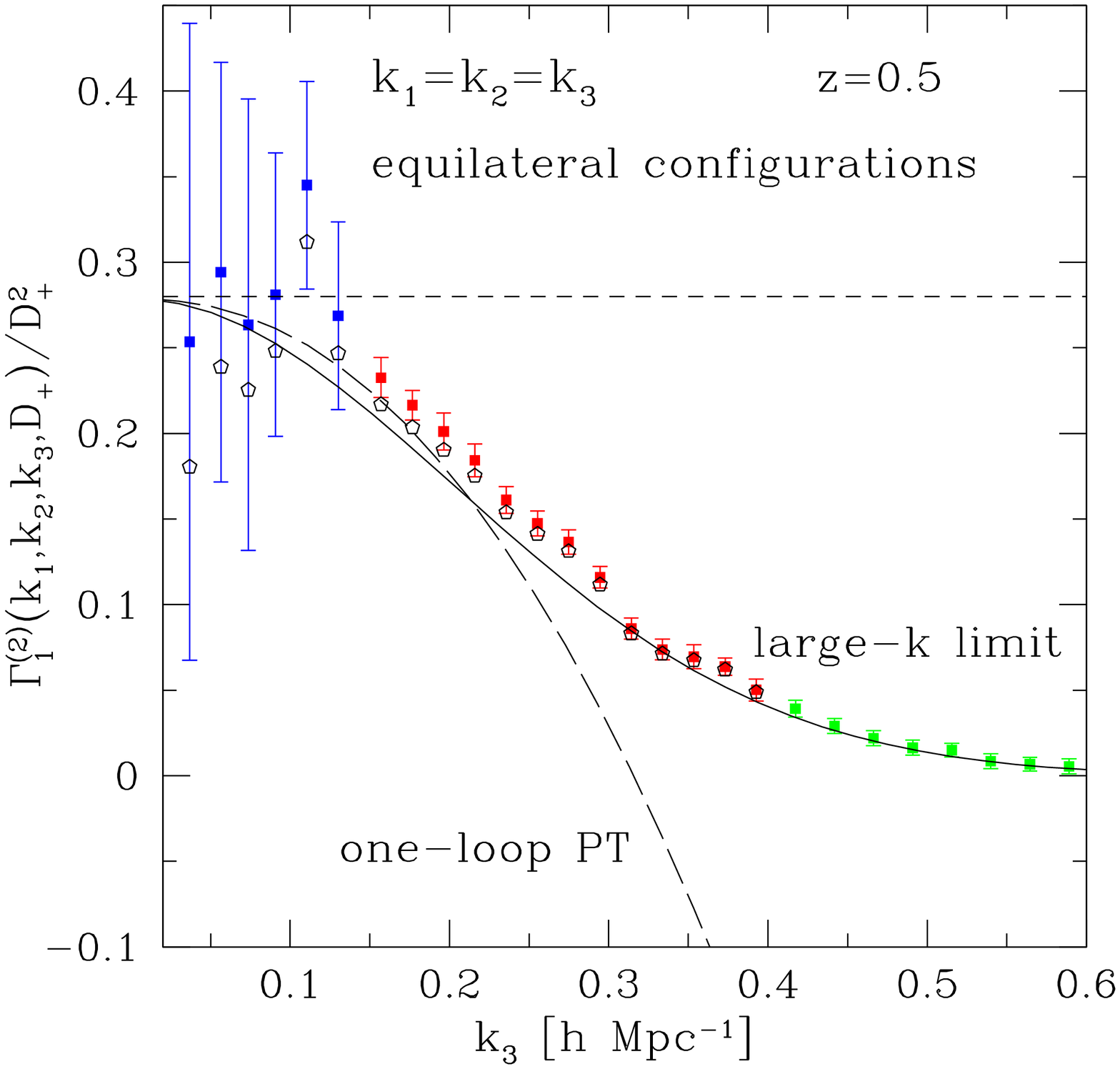} 
\includegraphics[width=0.4\textwidth]{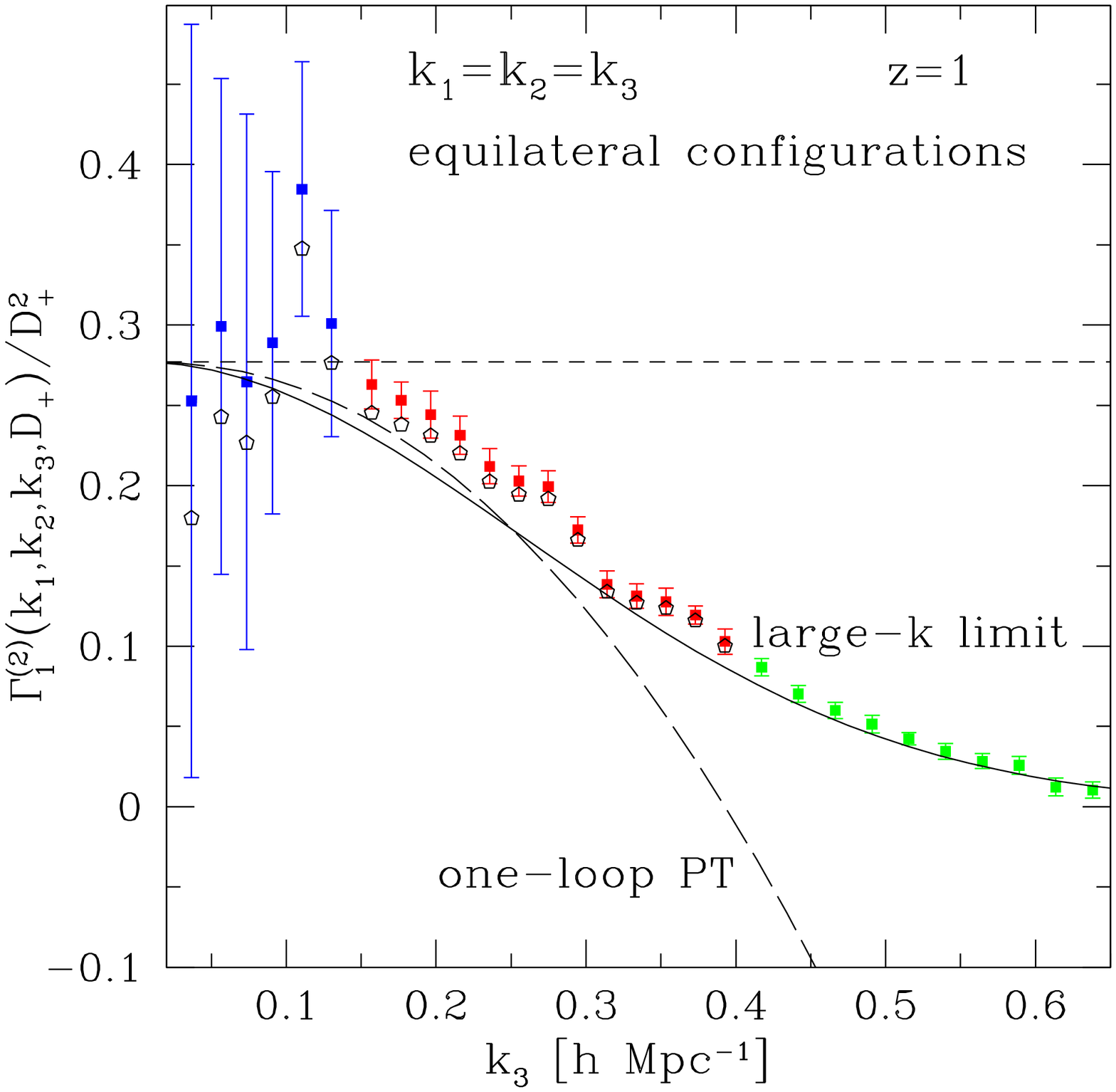}
\caption{Measurements of the three-point density propagator $\Gamma_1^{(2)}$ against analytical predictions for different triangle configurations and redshifts. Top Left Panel:  equilateral configurations at $z=0$. Top Right Panel: colinear configurations at $z=0$. Bottom Panels: equilateral configurations at $z=0.5$ (Left) and $z=1$ (Right). In each panel the long dashed line corresponds to the one-loop prediction given in Eqs.~(\ref{oneloopequilateral},\ref{oneloopcolinear}) while the solid line is the high-$k$ limit derived in Eq.~(\ref{G2Loops2}). At large scales the measurements reach the tree-level result well known from standard perturbation theory (horizontal short dashed line), given by Eq.~(\ref{Gamma1tree}). The empty circles in the panels of equilateral configurations are the measurements after the correction due to the estimator bias as described in the text. In addition, the sudden change in the size of error bars as a function of scale is due to the change in binning, see text for details.}
\label{F2measu}
\end{center}
\end{figure*}

Figure~\ref{F2measu} shows the density three-point propagator $\Gamma^{(2)}_1$ for equilateral configurations (top left panel for $z=0$ and bottom panels for $z=0.5,1$, as labeled) and colinear configurations (top right panel, $z=0$). In each panel the symbols with error bars correspond to the measurements in the N-body simulations, the long dashed line represents the one-loop result from Eqs.~(\ref{oneloopequilateral},\ref{oneloopcolinear}) and the solid line the large-$k$ limit from Eq.~(\ref{G2Loops2}). In all cases the errors displayed are for the mean of the ensemble obtained from the scatter among the $50$ realizations. For simplicity we only show the more familiar configurations, corresponding to colinear and equilateral triangles, although we have measured all configurations. 

The comparison in Fig.~\ref{F2measu} shows overall that both large-$k$ and low-$k$ regimes are well modeled by the analytic results for different triangle configurations and redshifts. This establishes a good basis from which a matching ansatz can be constructed to describe also intermediate scales, as done in \cite{2006PhRvD..73f3520C} for the two-point propagator.

Let us now point out some limitations in the measurements that help understand better the significance of the level of agreement seen in Fig.~\ref{F2measu}. The asymptotic limit to the tree level value at large scales is difficult to measure since few configurations of a given shape are available at low $k$, which gives rise to relatively large cosmic variance. To partially mitigate this one is tempted to increase the bin size in Fourier space. However, a larger bin size introduces a mixing of triangles of slightly different shape, which have different $\Gamma^{(2)}_1$. For equilateral triangles, increasing the bin size introduces slightly off equilateral triangles which bias the result to higher values as we show below, while the opposite is true for colinear configurations. Therefore a trade-off takes place between reducing the bin size to obtain unbiased results but at the expense of increasing the cosmic variance. We choose a bin size $\delta k=0.005 \kvecMpc=k_{\rm F} $ (the fundamental mode of the simulation volume) for $k \lesssim 0.15\kvecMpc$, $2k_{\rm F}$ that up to  $k \lesssim 0.4\kvecMpc$ and $4k_{\rm F}$ for smaller scales where the impact from offset configurations is negligible. This is the reason why the error bars diminish by a factor of a few at e.g. $k \sim 0.15 h^{-1}{\rm Mpc}$ for equilateral triangles. For colinear triangles, the change in binning happens at slightly different scales, evident from looking at the top right panel in Fig.~\ref{F2measu}.

The bias of the $\Gamma^{(2)}$ estimator can be understood and quantified using the tree level results as follows. Since $\vk_3=\vk_1+\vk_2$, equilateral triangles have an angle of $120^\circ$ between $\vk_1$ and $\vk_2$. Configurations slightly off this will give a tree level result, from Eq.~(\ref{Gamma1tree}), equal to
\begin{equation}
{1\over D_+^2}\ \Gamma^{(2)}_{1,{\rm tree}}(k,k,\cos(120^\circ+\delta))\approx\frac{2}{7}-\frac{5\sqrt{3}}{14}\delta,
\end{equation}
where $2/7$ is the ``exactly'' equilateral value. The off-equilateral configurations with larger $k_3$ have more statistical weight (more Fourier modes), and they are represented by negative $\delta$. Therefore a bias towards higher values of $\Gamma_1^{(2)}$ is introduced by increasing the bin size. Conversely, the slightly off-colinear configurations would bias the measurement below the exactly colinear case because 

\beq
{1\over D_+^2}\ \Gamma^{(2)}_{1,{\rm tree}}(k,k,\cos(0^\circ+\delta)) \approx 2-\frac{11}{14}\delta^2,
\eeq
resulting in a negative estimator bias. 

The empty circles in Fig.~\ref{F2measu} correspond to the measurements ``corrected'' for the estimator bias assuming that the configuration dependence of $\Gamma_1^{(2)}$, slightly away from the exactly colinear or equilateral configurations, is given by the tree-level expressions above (with $\delta\sim k_F / k $). For equilateral configurations the agreement with the one-loop prediction at intermediate scales is improved, particularly at $z=0.5$ and $1$. For colinear configurations the correction is negligible.

These estimates assume tree-level perturbation theory, when loop corrections are included the situation is expected to changed somewhat. As it is well known in the bispectrum case~\cite{1998ApJ...496..586S}, at low redshift the dependence on configuration is suppressed, so the estimator bias is expected to be larger at higher redshift where mixing different triangles leads to more bias. This helps explain why at higher redshift (bottom right panel in Fig.~\ref{F2measu}), the intermediate scale measurements done with bin size $2k_{\rm F}$ are more above the predictions than at low redshift. 


Finally, it is easy to show that the size of error bars on $\Gamma^{(2)}_a$ scales as $D_+^{-1}$, as for the bispectrum~\cite{1998ApJ...496..586S}. This is nicely recovered in Fig.~\ref{F2measu} when one compares $z=1,0.5$ and $0$.

\section{Applications to the Power Spectrum and Bispectrum}
\label{recon}

\subsection{Reconstructing the Power Spectrum from Multi-Point Propagators}
\label{recP}

\begin{figure}
\centerline{\epsfig {figure=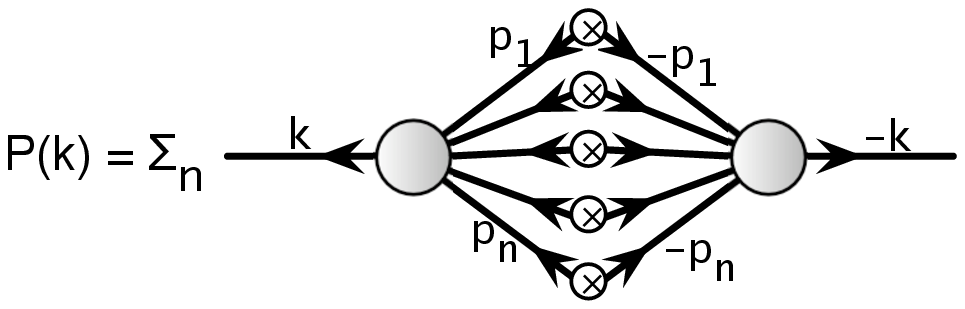,width=8cm}}
\caption{Reconstruction of the power spectrum from multi-point propagators, where a large shaded circle represents all possible loops that enter into the fully nonlinear multi-point propagator  (see Fig.~\ref{Gamma5} for its perturbative expansion). The crossed circles represent initial power spectra. The sum runs over the number of internal connecting lines, i.e. the number of such circles. It is important to note that each term in this sum is positive.}
\label{SpectreRec}
\end{figure}

We now discuss the relationship between multi-point propagators and nonlinear corrections to the power spectrum, showing that the power spectrum $P_{ab}$ can be reconstructed by gluing together $\Gamma^{(p)}$ contributions, as briefly mentioned in Eq.~(\ref{gammaexpansion}) above. 

A formal expression for $P_{ab}(k)$ can be written using the expansion in Eq.~(\ref{PsiExpansion}) as,
\begin{eqnarray}
\Dirac(\vk_{1}+\vk_{2})P_{ab}(k_{1},s)
&\equiv&\mg\Psi_{a}(\vk_{1},s)\Psi_{b}(\vk_{2},s)\md\nonumber\\
&=&\sum_{n_{1},n_{2}}\mg\Psi^{(n_{1})}_{a}(\vk_{1},s)\Psi^{(n_{2})}_{b}(\vk_{2},s)\md \nonumber
\\
\label{PkExpansion1}
\end{eqnarray}
Then, for a given choice of indices $n_{1}$ and $n_{2}$ one has to compute the ensemble average of $n_{1}+n_{2}$ factors $\phi(\vq_{i})$, following the field expansion in Eq.~(\ref{mFndef}). 

Since we assume Gaussian initial conditions, this joint ensemble average is then a sum of various terms, each of them the product of two-point correlators (Wick's theorem). Each of these terms can be labelled with indices $r,s,t$, where $r$ is the number of connected pairs
within the first $n_{1}$ fields $\phi_{c_{i}}(\vq_{i})$, $s$  is the number of connected pairs
within the last $n_{2}$ fields $\phi_{d_{i}}(\vq'_{i})$, and $t$ is the number of ``mixed''  pairs connecting
fields from the first $n_{1}$ and the last $n_{2}$. Obviously one has $n_{1}=2r+t$ and
$n_{2}=2s+t$.  Let us define 
$\mg \phi_{c_{1}}(\vq_{1})\dots\phi_{c_{2r+t}}(\vq_{2r+t});
\phi_{d_{1}}(\vq'_{1})\dots\phi_{d_{2s+t}}(\vq'_{2s+t})\md_{r,s,t}$
as the subset of the terms appearing in $
\mg \phi_{c_{1}}(\vq_{1})\dots\phi_{c_{2r+t}}(\vq_{2r+t})
\phi_{d_{1}}(\vq'_{1})\dots\phi_{d_{2s+t}}(\vq'_{2s+t})\md_c$
that correspond to a given $r,s,t$ triplet. 
\begin{widetext}
Then the discrete sum in (\ref{PkExpansion1}) can be written as a sum over
$r$, $s$ and $t$ as,
\begin{eqnarray}
\mg\Psi_{a}(\vk_{1})\Psi_{b}(\vk_{2})\md=
\sum_{r,s,t}
 \int\dd^3\vq_{1}\dots\dd^3\vq_{2r+t}\,\dd^3\vq'_{1}\dots\dd^3\vq'_{2s+t}\ 
\Dirac(\vk_{1}-\vq_{1\dots 2r+t})\ \Dirac(\vk_{2}-\vq'_{1\dots 2s+t})\nonumber\\
\times{\mF}_{ac_{1}\dots c_{2r+t}}^{(2r+t)}(\vq_{1},\dots ,\vq_{2r+t};\eta)\ \!\!
{\mF}_{bd_{1}\dots d_{2s+t}}^{(2s+t)}(\vq'_{1},\dots ,\vq'_{2s+t};\eta)
 \mg \phi_{c_{1}}(\vq_{1})\dots\phi_{c_{2r+t}}(\vq_{2r+t});
\phi_{d_{1}}(\vq'_{1})\dots\phi_{d_{2s+t}}(\vq'_{2s+t})\md_{r,s,t}
\label{PkExpansion2}
\end{eqnarray}
where we have omitted the time variable $\eta$ for clarity. This expression can be further written as,
\begin{eqnarray}
\mg\Psi_{a}(\vk_{1})\Psi_{b}(\vk_{2})\md=
\sum_{r,s,t}
t! (2r-1)!! (2s-1)!! \binom{2r+t}{t} \binom{2s+t}{t} 
\int\dd^3\vq_{1}\dots\dd^3\vq_{r}\,\dd^3\vq'_{1}\dots\dd^3\vq'_{s}\,\dd^3\vq''_{1}\dots\dd^3\vq''_{t}\nonumber\\
\times
\bar{\mF}_{a}^{(2r+t)}\left(\vq''_{1},\dots
,\vq''_{t},\vq_{1},-\vq_{1}\dots,\vq_{r},-\vq_{r};\eta\right)
\bar{\mF}_{b}^{(2s+t)}\left(-\vq''_{1},\dots
,-\vq''_{t},\vq'_{1},-\vq'_{1}\dots,\vq'_{s},-\vq'_{s}
;\eta\right)\nonumber\\
\times
P_{0}(q_{1})\dots P_{0}(q_{r})\ 
P_{0}(q'_{1})\dots P_{0}(q'_{s})\ 
P_{0}(q''_{1})\dots P_{0}(q''_{t})
\label{noname}
\end{eqnarray}
where $\bar{\mF}_{a}^{(n)}\equiv\mF_{a c_{1}\dots c_{n}}^{(n)}u_{c_{1}}\dots u_{c_{n}}$ and $P_0$ is the initial power spectrum from Eqs.~(\ref{gmic},\ref{Spectra}). Remarkably, one can recognize in this expression a sum of
products of $\Gamma^{(t)}$ functions, see Eq.~(\ref{GammaabcExp2}), leading to the formal expression,
\begin{eqnarray}
P_{ab}(\vk,\eta)=
\sum_{t} t! 
\int\dd^3\vq_{1}\dots\dd^3\vq_{t} \,
\Dirac(\vk-\vq_{1\dots t})\,
\Gamma^{(t)}_{a}\left(\vq_{1},\dots,\vq_{t};\eta\right)
\Gamma^{(t)}_{b}\left(\vq_{1},\dots,\vq_{t};\eta\right)
P_{0}(q_{1})\dots P_{0}(q_{t}),\label{PkExpansion}
\end{eqnarray}
\end{widetext}
where we have introduced the shorthand notation,
\begin{equation}
\Gamma^{(t)}_a(\vq_1,\ldots,\vq_t)=\Gamma^{(t)}_{a c_1 \ldots c_t}(\vq_1,\ldots,\vq_t)\,u_{c_1} \ldots u_{c_t},
\end{equation}
and used the following property,
\begin{equation}
\Gamma^{(n)}_{ab_{1}\dots b_{n}}\left(\vk_{1},\dots,\vk_{n}\right)=
\Gamma^{(n)}_{ab_{1}\dots b_{n}}\left(-\vk_{1},\dots,-\vk_{n}\right) \nonumber 
\label{parityInvar}\end{equation}

Thus, we have shown that the power spectrum can be written as a summation over contracted $n-$point propagators. This sum is diagrammatically represented in Fig.~\ref{SpectreRec}, and corresponds to a contraction of $\Gamma^{(n)}$ diagrams when the
incoming lines are glued together to form initial power spectra. The combinatorial factor $t!$ in front of Eq.~(\ref{PkExpansion}) is the number of ways of contracting two $(t+1)$-point propagators. 

It is interesting to note that for $a=b$, the result is a sum of positive terms. For example, for the density spectrum ($a=b=1$) up to one-loop corrections we have,

\beqa
P(k) &=& [\Gamma^{(1)}(k)]^2\, P_0(k) \nonumber \\
&+& 2\int d^3q\, [\Gamma^{(2)}(\vk-\vq,\vq)]^2  P_0(|\vk-\vq|)P_0(q). \nonumber \\
\label{P1loop}
\eeqa

This resummation scheme is then very attractive in practice, because, unlike standard PT, the subsequent contributions add constructively and correspond to a sum over ``bumps" peaked at increasingly higher $k$'s. We encounter here a similar situation as the one presented in the original formulation of RPT~\cite{2006PhRvD..73f3520C,2006PhRvD..73f3519C}, although the resummation scheme differs in detail. For example, here it is important to note that we are led to consider an infinite number of objects, the $n-$point propagators,  a property at variance with what a renormalization group approach suggests~\footnote{In particular, it seems unlikely that one could  obtain a local vertex renormalization, e.g. a renormalization flow in which $g \to G$ and $\gamma \to \tilde\gamma$ where the time dependence of $\Gamma$ would be only that of the time of interaction. The reason is that the results obtained here cannot simply be recap in such a form in the large-$k$ limit. If one wants to stick to such a renormalisation procedure, it would mean that the class of diagrams we have been able to resum will not be included.}. This is unlike in~\cite{2006PhRvD..73f3520C,2006PhRvD..73f3519C} where one deals with a few objects, e.g. the two-point propagator and the vertex (in this regard, note that Eq.~(\ref{PkExpansion}) automatically incorporates vertex renormalization). We have seen though that the multi-point propagators share common, and useful, properties. In particular, their expression for the high-$k$ limit ($k\sigma_v \gg 1$) provide a useful approximation to compute multi-loop corrections to the power spectrum, i.e. the mode-coupling power spectrum in Eq.~(\ref{PRPT}). 

We finalize this section by noting that the relation in Eq.~(\ref{PkExpansion}) can be also obtained by considering an expansion of the field $\Psi_a$ in terms of generalized Wiener-Hermite functionals of the initial fields $\phi_a$, see~\cite{1995ApJS..101....1M}. The multi-point propagators defined in Eq.~(\ref{GammaAllDef}) would then correspond to the kernel coefficients of this expansion. In ~\cite{1995ApJS..101....1M} it is also introduced the formal expansion to address the more general case of non Gaussian initial conditions. Note however that this approach does not provide a prescription to compute these kernels (e.g. a resummation scheme such as the one derived in Sec.\ref{largekGammap} or in~\cite{2006PhRvD..73f3519C} for $\Gamma^{(2)}$). 

\subsection{Reconstructing the Bispectrum from Multi-Point Propagators}
\label{recB}

\begin{figure}
\centerline{\epsfig {figure=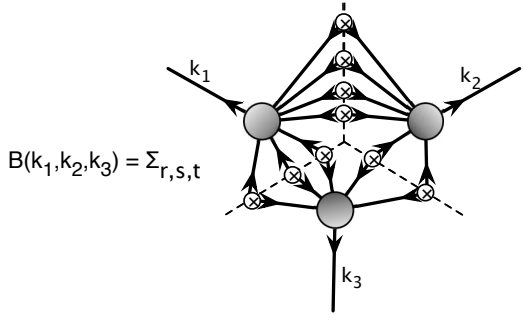,width=7cm}}
\caption{Reconstruction of the bispectrum from multi-point propagators. The crossed circles represent initial power spectra. The sum in Eq.~(\ref{BkExpansion}) runs over the number of connecting lines between each of the emerging modes, e.g. that cross each of the dashed half lines.}
\label{BiSpectreRes}
\end{figure}

We now extend these results to the bispectrum. Generalizing Eq.~(\ref{PkExpansion}) it is easy to show that the bispectrum can be formally written as,

\begin{widetext}
\begin{eqnarray}
\mg\Psi_{a}(\vk_{1})\Psi_{b}(\vk_{2})\Psi_{c}(\vk_{2})\md&=&
\sum_{r,s,t} \binom{r+s}{r}\binom{s+t}{s}\binom{t+r}{t} r! s! t! 
\int\dd^3\vq_{1}\dots\dd^3\vq_{r}
\ \dd^3\vq'_{1}\dots\dd^3\vq'_{s}
\ \dd^3\vq''_{1}\dots\dd^3\vq''_{t}\ \nonumber\\
&&\,\times
\Dirac(\vk_{1}-\vq_{1\dots r}-\vq'_{1\dots s})\ 
\Dirac(\vk_{2}+\vq'_{1\dots s}-\vq''_{1\dots t})\ 
\Dirac(\vk_{3}+\vq''_{1\dots t}+\vq_{1\dots r})   \nonumber \\
&&\,\times
\Gamma^{(r+s)}_a\left(\vq_{1},\dots,\vq_{r},\vq'_{1},\dots,\vq'_{s}\right) 
\Gamma^{(s+t)}_b\left(-\vq'_{1},\dots,-\vq'_{s},\vq''_{1},\dots,\vq''_{t}\right)\nonumber\\
&&\times\,
\Gamma^{(t+r)}_c\left(-\vq''_{1},\dots,-\vq''_{t},-\vq_{1},\dots,-\vq_{r}\right)
P_{0}(q_{1})\dots P_{0}(q_{r})\ P_{0}(q'_{1})\dots P_{0}(q'_{s})\ P_{0}(q''_{1})\dots
P_{0}(q''_{t}). \nonumber \\
\label{BkExpansion}
\end{eqnarray}
This sum is diagrammatically represented in Fig.~\ref{BiSpectreRes}. We see that it runs over the number of lines that connect each side of the diagram (with the constraint that at most one of the indices $r$, $s$ or $t$ is zero, otherwise we would have a disconnected diagram). The leading order (tree) contribution is then obtained for $r=s=1$, $t=0$ (plus cyclic permutations), up to one-loop corrections (in square brackets) we have 

\beqa
B(k_1,k_2,k_3)&=& 2\, \Gamma^{(2)}(\vk_1,\vk_2)\, \Gamma^{(1)}(k_1)\, \Gamma^{(1)}(k_2)\, P_0(k_1)P_0(k_2)+ {\rm cyc.} \nonumber \\
 &+&
\Big[ 8 \int d^3q\,  \Gamma^{(2)}(\vk_1-\vq,\vq) \Gamma^{(2)}(\vk_2+\vq,-\vq) \Gamma^{(2)}(\vq-\vk_1,-\vk_2-\vq) P_0(|\vk_1-\vq|) P_0(|\vk_2+\vq|) P_0(q)  \nonumber \\
 &+&
6  \int d^3q\,  \Gamma^{(3)}(-\vk_3,-\vk_2+\vq,-\vq)  \Gamma^{(2)}(\vk_2-\vq,\vq) \Gamma^{(1)}(\vk_3) P_0(|\vk_2-\vq|)P_0(q)P_0(k_3)+ {\rm cyc.}
\Big].
\label{B1loop}
\eeqa
\end{widetext}
Note that having resummed the multi-point propagators means that many of the one-loop corrections in standard PT are already encoded in $\Gamma^{(p)}$ and thus the number of one-loop diagrams is reduced. For the power spectrum we have one instead of two diagrams, for the bispectrum we have two instead of the four in standard PT~\cite{1998ApJ...496..586S}. 

It is useful to compare the structure of Eqs.~(\ref{P1loop}) and~(\ref{B1loop}). We see that the one-loop corrections to the power spectrum  depend on the initial power spectrum $P_0$ through a convolution with the three-point propagator $\Gamma^{(2)}$, which determines the large-scale (tree-level) bispectrum. The two-loop correction to the power spectrum involves a similar convolution with $\Gamma^{(3)}$, which determines the large-scale trispectrum, and contributes to the one-loop bispectrum. This pattern continues to higher orders, demonstrating that in order to extract the most information about the initial power spectrum $P_0$, it is advantageous to simultaneously measure the power spectrum and higher-order spectra at large scales and include these relationships when doing cosmological parameter estimation.

As a preliminary application of these results, we compute the reduced bispectrum $Q$ defined by

\beq 
Q = \frac{B(k_1,k_2,k_3)}{P(k_1)P(k_2)+P(k_2)P(k_3)+P(k_3)P(k_1)},
\label{Qdef}
\eeq
where we use one-loop results for both the power spectrum and bispectrum from Eq.~(\ref{P1loop}) and Eq.~(\ref{B1loop}), respectively. Since we don't yet have a full prescription for the multi-point propagators valid at all scales, we use their high-$k$ limit expressions, Eq.~(\ref{Gammapq1loops}) modified as follows,

\beq
\Gamma^{(p)}= \frac{\Gamma^{(1)}(k_p)}{ \Gamma^{(1)}_{{\rm tree}}(k_p)}\ 
\Gamma^{(p)}_{{\rm tree}},
\label{GammaTrucho}
\eeq
which reduces to Eq.~(\ref{Gammapq1loops}) in the high-$k$ limit, and approximately incorporates (through the full scale-dependence of the two-point propagator $\Gamma^{(1)}$ found in~\cite{2006PhRvD..73f3520C}) the fact that at low-$k$ propagators decay slower than their high-$k$ limit indicates, as shown in Fig.~\ref{F2measu} by comparing the one-loop to the high-$k$ expressions at low-$k$. 

\begin{figure}
\centerline{\epsfig {figure=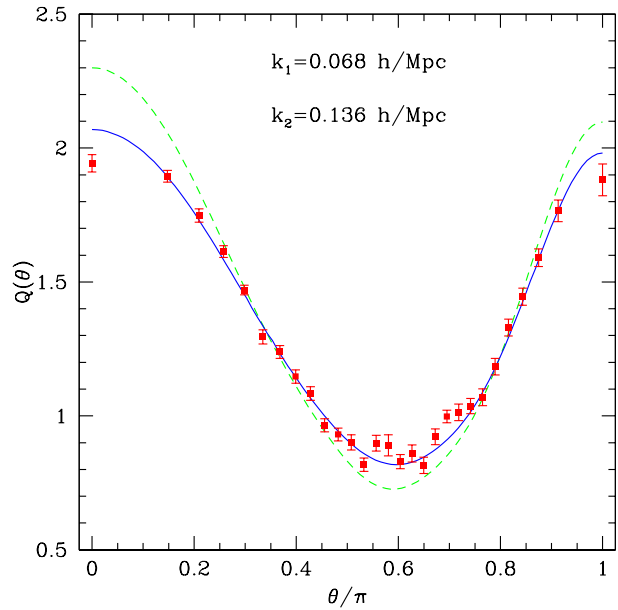,width=8.5cm}}
\caption{The reduced bispectrum $Q$ at $z=0$ in tree-level PT (dashed line), one-loop RPT (solid line) and measured in numerical simulations (symbols with error bars) as a function of the angle between $\vk_1$ and $\vk_2$.}
\label{QRPT}
\end{figure}

Figure~\ref{QRPT} shows the result of this calculation (solid lines) compared to measurements of the bispectrum in our numerical simulations (symbols with error bars). Because we only compute up to one-loop corrections, the range of scales available to test our predictions is limited. At low-$k$, the deviations from tree-level PT are small, and error bars increase. At high-$k$ two-loop corrections become important (see e.g. Fig.~3 in~\cite{2008PhRvD..77b3533C} for the case of the power spectrum). We concentrate on $z=0$ where the deviations from tree-level are largest and the application of standard PT is least successful. The scales shown in Fig.~\ref{QRPT} are those where the deviations from tree-level are largest while we seem to be still within the validity of the one-loop approximation. We see that the level of agreement with simulations is very encouraging despite the approximate nature of Eq.~(\ref{GammaTrucho}), correctly describing the loss of dependence of $Q$ on triangle shape. More work is needed to explore the robustness of these predictions, but it is clear that the results found in this paper on multi-point propagators will be useful for a variety of statistics. A  detailed discussion of the bispectrum will be presented elsewhere.

\section{Conclusions}

We generalized the notion of two-point propagator to arbitrary multi-point propagators. We derived the one-loop corrections to the three-point propagator, and were able to resum the dominant class of diagrams contributing to multi-point propagators in the large-$k$ limit, extending what was derived before  for the two-point propagator in~\cite{2006PhRvD..73f3520C}. Our results show that multi-point propagators decay with the same rate as the two-point propagator in the large-$k$ limit and are proportional to their tree-level values. We have extended the algorithm presented in~\cite{2006PhRvD..73f3520C} to measure the three-point propagator in numerical simulations, and showed that our high-$k$ limit resummation agrees well with both two and three-point propagators at nonlinear scales.

Furthermore, we showed that the multi-point propagators can be considered building blocks out of which one can construct arbitrary $n-$point spectra. In addition, nonlinear corrections to the power spectrum depend on the initial spectrum through convolutions with multi-point propagators, which in turn determine the large-scale (tree-level) polyspectra. This gives a direct physical connection between nonlinear corrections to the power spectrum at small scales and higher-order correlations at large scales.

We used our results on the resummation of multi-point propagators to reconstruct the power spectrum and bispectrum up to one-loop level, and showed that the predictions for the reduced bispectrum agree well with measurements in numerical simulations. This is encouraging news for the calculation of higher-order correlation functions in renormalized perturbation theories.

\begin{acknowledgments}

This work was supported in part by the French Programme National de Cosmology and by the French Agence National de la Recherche under grant BLAN07-1-212615. MC acknowledges from support from Spanish Ministerio de Ciencia y Tecnologia (MEC), project AYA2006-06341 and the Juan de la Cierva MEC program. RS is partially supported by NSF AST-0607747 and NASA NNG06GH21G.

\end{acknowledgments}

\hspace{2cm}
\newpage
\appendix

\section{Time-Ordering and Equation~(\ref{G2Loops2})}
\label{TOrd}

Here we provide another derivation of the resummation of the three-point propagator that explicitly takes into account time-ordering. An $n$-loop diagram is specified by the number of vertices $n_i$ ($i=1,2,3$) along the principal lines with momenta $k_i$, i.e. $2n=n_1+n_2+n_3$, $n_1=2p_{11}+p_{12}+p_{13}$, $n_2=2p_{22}+p_{12}+p_{23}$, $n_3=2p_{33}+p_{13}+p_{23}$, where $p_{ij}$ are defined as in section~\ref{Gamma2Res} (note that $n=\sum_{i\leq j}p_{ij}$). Here we start from scratch, i.e. to construct such an $n$-loop diagram we start from the principal tree, to which we attach $n_i$ branches to principal lines of momentum $k_i$ each with its initial condition (as in Fig.~\ref{DiagramPsiExpansion}) and we proceed to take expectation values. 

Consider one of the principal lines at a time, say (3) carrying momentum $k_3$. There are $n_3$ time-ordered vertices along this line, and its symmetry factor is $2^{n_3}$ because branching is asymmetric (which leads to Eq.~(\ref{sijexp}) when all three principal lines are included). Out of these $n_3$ branches one must select $2p_{33}$ for averaging among themselves (thus making $p_{33}$ loops), out of the remaining $p_{13}+p_{23}$ one must select $p_{13}$ for averaging with $p_{13}$ branches attached to the principal line (1), thus creating $p_{13}$ loops, and the remaining $p_{23}$ average with $p_{23}$ branches attached to (2). The number of ways of doing this is,
\beq
{n_3 \choose 2p_{33}} { p_{13}+p_{23} \choose p_{13}} = \frac{n_3!}{(2p_{33})! \, p_{13}! \,p_{23}!}
\label{CombFest}
\eeq
and the analogous combinatorial factors also apply to the other 2 principal lines. The process of taking expectation values introduces the following factors: $(2p_{ii}-1)!! $ is the number of ways of averaging pairs (Gaussian fields) out of $2p_{ii}$ initial conditions, while there are $p_{ij}!$ ways of connecting $p_{ij}$  branches ($i\neq j$) coming out of principal lines ($i$) and ($j$). This leads to the combinatorial factor,

\beq
(2p_{11}-1)!! \, (2p_{22}-1)!! \, (2p_{33}-1)!!\,   p_{12}! \, p_{13}! \,p_{23}!
\label{CombFest2}
\eeq
Putting together this with Eq.~(\ref{CombFest}) and its two permutations for the two other principal lines (and the symmetry factors for each line) we have a total overall factor of

\beq
\frac{2^{n_1+n_2+n_3}\ n_1! \ n_2! \ n_3! }{2^{p_{11}+p_{22}+p_{33}} \prod_{i\leq j} p_{ij}!   }
\label{CombTot}
\eeq
Finally the time-ordered integrals give, e.g. for principal line (3), 

\beq
\int_{s'}^{s} ds_1 {\rm e}^{s_1} \int_{s'}^{s_1} ds_2 {\rm e}^{s_2} \ldots \int_{s'}^{s_{n_3-1}} ds_{n_3} {\rm e}^{s_{n_3}} =
{ ({\rm e}^s-{\rm e}^{s'})^{n_3}\over n_3!},
\label{Tord}
\eeq
where $s$ is the final time and the principal tree splits at time $s'$. For principal lines (1) and (2) one obtains analogous factors but with the corresponding time dependences, i.e. $({\rm e}^{s'}-1)$. Therefore, we see that the $n_i!$ factors in Eq.~(\ref{CombTot}) get cancelled by time-ordering (as expected from the discussion in section~\ref{Gamma2Res}, see in particular Fig.~\ref{Gamma2Sym}). Then, after momentum integration, which gives $-\sigma_v^2 k_i k_j / 4 $ for loops attached to principal lines ($i$) and ($j$), we can write this equation taking into account time ordering as follows, 

\begin{widetext}

\begin{eqnarray}
\Gamma^{(2)}_{abc,\{p_{ij}\}}&=&
\prod_{i} {(-k_{i}^2\sigma_v^2/2)^{p_{ii}} \over p_{ii}!}\ \prod_{i<j}
{ \left(-\vk_{i}.\vk_{j}\sigma_v^2 \right)^{p_{ij}} \over p_{ij}!}
\int_{0}^{s}\dd s' g_{ad}(s-s')\gamma_{def}(\vk_{1},\vk_{2},\vk_{3})
g_{eb}(s')g_{fc}(s')\nonumber\\
&&\times \left(e^{s'}-1\right)^{2p_{11}+2p_{22}+2p_{12}+p_{13}+p_{23}}
\left(e^s-e^{s'}\right)^{2p_{33}+p_{13}+p_{23}},
\label{Gamma2new}
\end{eqnarray}
\end{widetext}
which when summed over the $p_{ij}$, and using that $\vk_3=\vk_1+\vk_2$, gives the tree-level value times a Gaussian factor, i.e. Eq.~(\ref{G2Loops2}).

\section{Resummation in Fourier space shells}
\label{ResFS}

The results presented in the main part of this paper rely on a direct resummation of multiloop contributions. In this appendix we present another method that makes use of the mathematical structure of the loop correction terms in the large-$k$ limit. The idea is to partition Fourier space into finite volume elements. Although there is no need to actually specify the partition we will be using, one simple way of constructing an explicit one is to decompose the volume into concentric shells that we will denote $i,j,\dots$, with the $i$-th shell corresponding to $q\in[\Lambda_{i-1},\Lambda_{i}]$ and being of width $\delta\Lambda_{i}$. That means that the integration over Fourier modes,  $\int\dd\vq^3$,  becomes $4\pi \sum_{i} \Lambda_{i}^2 \delta\Lambda_{i}$ assuming that the integrand depends on the norm of $\vq$ only. Let us now explore the loop contributions for the two-point propagator $G_{ab}$ in this framework.

According to the results presented in section~\ref{largekGammap}, see e.g. Eq.~(\ref{G2Loopij1}), the contribution to the one-loop correction coming from shell $i$ to
$G_{ab}(k)$, $\delta^{i} G_{ab}(k)$, reads in the large-$k$ limit
\begin{equation}
\delta^{i} G_{ab}(k)=g_{ab}(k)\ \mL(k,\Lambda_{i})\delta\Lambda_{i},
\end{equation}
where we used the following definitions,
\begin{eqnarray}
\mL(s_2,s_1,k,\Lambda)&=&-\frac{2\pi{k^2}}{3}e^{s_{1}} e^{s_{2}} \,P_{0}(\Lambda)
\label{mLdef}\\
\mL(k,\Lambda)&=&\int_{0}^{s}\dd s_{1}\int_{0}^{s} \dd s_{2}\ \mL(s_{2},s_{1},k,\Lambda)\nonumber\\
&=&
-\frac{2\pi{k^2}}{3}(e^{s}-1)^2 \,P_{0}(\Lambda).
\label{mLdef2}
\end{eqnarray}
Multiloop corrective terms exhibit then a truly remarkable property. Let us start with
a two-loop contribution. In all of these calculations the linear propagator factors along
the principal line can be factorized in the large-$k$ limit, due to the property that $g_{ab}(s-s')g_{bc}(s')=g_{ac}(s)$.
Thus, we simply have for the two-loop contribution to $G_{ab}$, $\delta^{i,j}G_{ab}$,
\begin{eqnarray}
\delta^{i,j} G_{ab}(k)=g_{ab}\ \mL(k,\Lambda_{i})\delta\Lambda_{i}\
\mL(k,\Lambda_{j})\delta\Lambda_{j}.
\end{eqnarray}
This factorization property can be generalized to any order. As a result, for a collection of loops
whose indices are in set $\mathcal{I}$, the multi-loop correction term, $\delta^{\mathcal{I}}G_{ab}$, reads
\begin{equation}
\delta^{\mathcal{I}} G_{ab}(k)=g_{ab}\prod_{i\in \mathcal{I}}\left[ \mL(k,\Lambda_{i})
\delta\Lambda_{i}\right].
\label{DiffGabFactProp}
\end{equation}

We can then resum all these contributions, e.g. resum $\delta^{\mathcal{I}} G_{ab}$ for relevant
sets $\mathcal{I}$ of shell indices. Let us denote $G_{ab,\Lambda_{i}}(k)$ the resummed value
of the propagator when all contributions of shells up to $i$  have been taken into account, that
is, assuming there is a cutoff for wavemode $q$ at $q=\Lambda_{i}$.

The result is formally given by a sum of terms
of the form $\delta^{\mathcal{I}} G_{ab}(k)$ where the sum runs over all possible subsets
$\mathcal{I}$ of
$\{1,\dots,i\}$, e.g.,
\begin{equation}
G_{ab,\Lambda_{i}}=\sum_{\mathcal{I}\subset \{1,\dots,i\} } \delta^{\mathcal{I}} G_{ab}(k),
\end{equation}
and when the width of each shell is made infinitely small. The subsets of $\{0,\dots,i\}$ can be
split into the subsets of $\{0,\dots,i-1\}$ and
the subsets of $\{0,\dots,i\}$ that contains $i$. As a result this sum can be rewritten,
\begin{eqnarray}
\sum_{\mathcal{I}\subset {1,\dots,i} } \delta^{\mathcal{I}} G_{ab}(k) &=&
\sum_{\mathcal{I}\subset{1,\dots,i-1} } \delta^{\mathcal{I}} G_{ab}(k)\nonumber\\
&+&\sum_{\mathcal{I}\subset{1,\dots,i-1} }\delta^{\mathcal{I}\cup\{i\}} G_{ab}(k).
\end{eqnarray}
Because of the factorization property, (\ref{DiffGabFactProp}), this decomposition then reads
\begin{equation}
G_{ab,\Lambda_{i}}= G_{ab,\Lambda_{i-1}}+G_{ab,\Lambda_{i-1}}\mL(k,\Lambda_{i})  \delta\Lambda_{i}.
\label{DiffGab2}
\end{equation}

In the continuous limit it leads to the equation,
\begin{equation}
\delta G_{ab}(k,\Lambda)=G_{ab}(k,\Lambda)\
\mL(k,\Lambda)\delta\Lambda.
\label{DiffGab}
\end{equation}
The solution of this ODE is Eq.~(\ref{GabHighk}) for the initial condition $G_{ab}(k,0)=g_{ab}(k)$.

This procedure can be simply extended to multi-point propagators.
Let us denote $\Gamma^{(2)}_{abc,\Lambda_{i}}$ the expression of $\Gamma^{(2)}_{abc}$ when modes above $\Lambda_{i}$ are cut out. In a construction similar to that of the two-point propagator, the structure of the loop correction is such that,
\begin{eqnarray}
\delta^{\mathcal{I}} \Gamma^{(2)}_{abc}(\vk_{1},\vk_{2},\vk_{3})&=&
\Gamma^{(2)}_{abc,{\rm tree}}(\vk_{1},\vk_{2},\vk_{3})\nonumber\\
&&\hspace{-2cm}\times\prod_{i\in \mathcal{I}}\left[
\mL(k_{3},\Lambda_{i}) \delta\Lambda_{i}\right].
\end{eqnarray}
It them implies that
\begin{eqnarray}
\Gamma^{(2)}_{abc,\Lambda_{i}}(\vk_{1},\vk_{2},\vk_{3})-\Gamma^{(2)}_{abc,\Lambda_{i-1}}(\vk_{1},\vk_{2},\vk_{3})
&=&\nonumber\\
\hspace{-2cm}\Gamma^{(2)}_{abc,\Lambda_{i}}(\vk_{1},\vk_{2},\vk_{3})\mL(k_{3},\Lambda_{i})
\delta\Lambda_{i}.
\end{eqnarray}
In the continuous limit it leads to a simple ODE that can straightforwardly
integrated into (\ref{G2Loops2}) taking into account the initial condition,
\begin{equation}
\Gamma^{(2)}_{abc,\Lambda=0}=\Gamma^{(2)}_{abc,{\rm tree}}.
\end{equation}

This construction extends to any multi-point propagator, or more precisely the procedure
extends to the computation of the loop corrections that apply to any  distinct tree
structure of tree order multi-point propagators.

\bibliography{GrandesStructures}

\end{document}